\documentclass[dvipdfmx,aps,prd,showpacs,showkeys,preprintnumbers,amsmath,amssymb,nofootinbib,10pt]{revtex4}
 \usepackage{eepic}
 \usepackage{ulem}
 \usepackage{tabularx}
\usepackage{float}
\usepackage{graphicx}
\usepackage{hyperref}
\usepackage{makecell}

\newcommand{\be}{\begin{equation}}
\newcommand{\ee}{\end{equation}}
\newcommand{\ba}{\begin{array}{c}}
\newcommand{\ea}{\end{array}}
\newcommand{\bqa}{\begin{eqnarray}}
\newcommand{\eqa}{\end{eqnarray}}
\newcommand{\bqaa}{\begin{eqnarray*}}
\newcommand{\eqaa}{\end{eqnarray*}}
\newcommand{\mL}{\mathcal{L}}
\newcommand{\cO}{\mathcal{O}}
\newcommand{\bra}{\langle}
\newcommand{\ket}{\rangle}
\newcommand{\nn}{\nonumber}

\begin{document}

\preprint{   IFT-UAM/CSIC-15-020  }
\preprint{   FTUAM-15-6  }

\title {
Scrutinizing the $\eta$-$\eta'$ mixing, masses and pseudoscalar decay constants \\
in the framework of $U(3)$ chiral effective field theory
}

\author{ Xu-Kun Guo$^a$}
\author{ Zhi-Hui~Guo$^{a,b}$}\thanks{ Corresponding author: zhguo@mail.hebtu.edu.cn }
\author{ Jos\'e Antonio~Oller$^c$}
\author{Juan~Jos\'e~Sanz-Cillero$^d$}
\affiliation{ ${}^a$  Department of Physics, Hebei Normal University, Shijiazhuang 050024, People's Republic of China
  \\ ${}^b$ State Key Laboratory of Theoretical Physics, Institute of
Theoretical Physics, CAS, Beijing 100190, People's Republic of China \\
 ${}^c$ Departamento de F\'isica, Universidad de Murcia,  E-30071 Murcia, Spain \\
 ${}^d$ Departamento de F\'\i sica Te\'orica and Instituto de F\'\i sica Te\'orica, IFT-UAM/CSIC, Universidad Aut\'onoma de Madrid, Cantoblanco, 28049 Madrid, Spain
}

\begin{abstract}
We study the $\eta$-$\eta'$ mixing up to next-to-next-to-leading-order in $U(3)$ chiral perturbation theory
in the light of recent lattice simulations and phenomenological inputs.
A general treatment for the $\eta$-$\eta'$ mixing at higher orders, with the higher-derivative, kinematic and mass mixing terms, is addressed.
The connections between the four mixing parameters in the two-mixing-angle scheme and the low energy constants in
the $U(3)$ chiral effective theory are provided both for the singlet-octet and the quark-flavor bases.
The axial-vector decay constants of pion and kaon are studied in the same
order and confronted with the lattice simulation data as well.
The quark-mass dependences of $m_\eta$, $m_{\eta'}$ and $m_K$ are found to be
well described at next-to-leading order.
Nonetheless, in order to simultaneously describe the lattice data
and phenomenological determinations for the properties of light pseudoscalars $\pi, K, \eta$ and $\eta'$,
the next-to-next-to-leading order study is essential.
Furthermore, the lattice and phenomenological inputs
are well reproduced for reasonable values of low the energy constants, compatible with previous bibliography.
\end{abstract}


\maketitle

\newpage
\tableofcontents

\section{Introduction}

The phenomenology of light flavor pseudoscalar mesons $\eta$ and $\eta'$ provides a valuable window on many important
nonperturbative features of Quantum Chromodynamics (QCD). It includes such important aspects as:
\begin{itemize}
\item{}
The spontaneous breaking of chiral symmetry, which gives rise to the appearance of the multiplet of light pseudoscalar mesons.
\item{}
The $U(1)_A$ anomaly of strong interactions, which gives mass to the singlet $\eta_0$ in $N_C=3$ QCD,
even in the chiral limit.
\item{}
The explicit $SU(3)$-flavor symmetry breaking, due to the splitting $m_s\neq \hat{m}$ between the strange and up/down quark masses
(the isospin limit, where $m_u=m_d=\hat{m}$ and the electromagnetic corrections are neglected, will be assumed all through the article).
\item{}
The $1/N_C$ expansion of QCD in the limit of large $N_C$, with $N_C$ the number of colors in QCD.
\end{itemize}
The interaction between the pseudo-Nambu-Goldstone bosons (pNGBs) $(\pi,K,\eta_8)$ from the spontaneous chiral symmetry breaking can be
systematically described through a low-energy effective field theory (EFT) based on $SU(3)_L\times SU(3)_R$
chiral symmetry, namely Chiral Perturbation Theory ($\chi$PT)~\cite{gasser8485}.
Following large--$N_C$ arguments~\cite{Nc},
this approach was later extended, incorporating the singlet  $\eta_0$ into a $U(3)$ $\chi$PT
Lagrangian~\cite{u3lo,leutwyler98npbsp,Kaiser:1998ds,kaiser00epjc,herrera97npb,herrera98plb}.
This combination of $\chi$PT and the $1/N_C$ expansion provides
a consistent framework which addresses all the previous issues.

More precisely, in this article we show that this large--$N_C$ $\chi$PT framework
yields an excellent description of the $\eta$ and $\eta'$ masses from
lattice simulations at different light-quark masses~\cite{hsc11prd,rbcukqcd10prl,ukqcd12prd,etm13pos,etm13prl}.
Constraints from phenomenological studies of $\rho,\, \omega,\, \phi,\, J/\psi$
decays~\cite{chen12prd,Chen:2014yta}
and kaon mass lattice simulations~\cite{rbcukqcd11prd,rbcukqcd13prd}
are compatible and easily accommodated in a joint fit.
The problems arise when one tries to also describe lattice simulations for $F_\pi$, $F_K$
and $F_K/F_\pi$~\cite{rbcukqcd11prd,rbcukqcd13prd,Durr:2010hr}.
Nevertheless, the issue of these observables in $\chi$PT is known and has been widely discussed in previous
bibliography~\cite{DescotesGenon:1999uh,Bijnens:2011tb,Bijnens:2014lea,Ecker:2013pba,Guo:2014yva}. It constitutes
a problem in its own and it is not the central goal of this article. It is discussed for sake of completeness and to show
its impact in a global fit.

The $\eta$ and $\eta'$ mesons not only attract much attention from the chiral community but
they have been also intensively scrutinized in lattice QCD simulations,
where enormous progresses have been recently made by different
groups~\cite{hsc11prd,rbcukqcd10prl,ukqcd12prd,etm13pos,etm13prl}.
Varying the light-quark masses $\hat{m}$ and $m_s$, both their masses and mixing angles have been extracted
in the range 200~MeV~$<m_\pi<$700~MeV. We will focus on the simulation points with $m_\pi<500$~MeV in the present work.
By observing the dependence of these observables with the light-quark masses we will
determine the $\chi$PT low energy constants (LECs) and further constrain the theoretical models.
At the practical level we have recast all $\hat{m}$ dependencies in terms of $m_\pi$ and study the observables
as functions of $m_\pi$.
The $\eta$ and $\eta'$ lattice simulations have not been thoroughly analyzed in the chiral framework yet
and it is the central goal of the present work.
However, the numerical uncertainties resulting from our analyses in this work
must be taken with a grain of salt as correlations
between the different lattice data points and other systematic errors are not considered here.

In addition to lattice QCD, there are also phenomenological studies of
the $\eta$ and $\eta'$ mixing, which has been extensively investigated in
radiative decays of light-flavor vector resonances $\rho,\omega,\phi$ and
$J/\psi \to VP, P\gamma$ processes~\cite{feldmann98prd,escribano05jhep,thomas07jhep,zhao08jpg,chen12prd,escribano14prd,Chen:2014yta,DeFazio:2000my,Schechter:1992iz}.
In these works, the modern two-mixing-angle scheme for the $\eta$ and $\eta'$ mesons, which was first advocated
in Refs.~\cite{leutwyler98npbsp,Kaiser:1998ds},
was employed to fit various experimental data.
The common methodology in these works is that the two-mixing-angle pattern for the $\eta$ and $\eta'$ is simply adopted to perform the
phenomenological discussion and the mixing parameters are then directly determined from data.
This is a bottom-up approach to address the $\eta$-$\eta'$ mixing problem and it is quite useful for the phenomenological analysis.
Contrary to the bottom-up method, it is also very interesting to study the $\eta$-$\eta'$ mixing from a top-down approach in which
one first constructs the relevant $\chi$PT Lagrangian and then calculates the $\eta$-$\eta'$ mixing pattern and parameters in terms of the
LECs. In this case, one can predict the $\eta$-$\eta'$ mixing parameters once the values of the unknown LECs are given.
The present work belongs to the latter category of top-down approaches.

Though the singlet $\eta_0$ meson, which is the main component of the physical $\eta'$ state, is not a pNGB due to
the strong $U(1)_A$ anomaly,
it can be formally introduced into $\chi$PT from the large-$N_C$ point of view.
The argument is that the quark loop induced $U(1)_A$ anomaly,
which is responsible for the large mass of the singlet $\eta_0$, is $1/N_C$ suppressed and hence the $\eta_0$
becomes the ninth pNGB in the large $N_C$ limit~\cite{ua1nc}. Based on this argument, the leading-order (LO)
effective Lagrangian for $U(3)$ $\chi$PT, which simultaneously includes the pNGB octet $\pi, K, \eta_8$ and the singlet $\eta_0$ as dynamical fields,
was formulated in Ref.~\cite{u3lo}. Later on, a full $\cO(p^4)$ $U(3)$ chiral Lagrangian was constructed in Ref.~\cite{herrera97npb}
and the discussion on the $\cO(p^6)$ unitary group chiral Lagrangian has been very recently completed in Ref.~\cite{Jiang:2014via}.
Subtle problems about the choice of suitable variables for the higher order $U(3)$ $\chi$PT Lagrangian in the large $N_C$ framework were analyzed
in Ref.~\cite{kaiser00epjc}.

The standard power counting employed in $SU(2)$ and $SU(3)$~$\chi$PT
in powers of the external momenta and quark masses~\cite{gasser8485},
is not valid any more in $U(3)$ $\chi$PT, due to the appearance of the large $\eta_0$ mass. However, since the singlet $\eta_0$ mass squared
behaves like $1/N_C$ in large $N_C$ limit, the $\eta_0$ mass can be harmonized with the other two expansion parameters if one assigns
the same counting
to $1/N_C$, the squared momenta $p^2$ and the light quark masses $m_q$. As a result of this,
in order to have a systematic power counting,
the combined expansions on momentum, light quark masses and $1/N_C$  are mandatory in $U(3)$ $\chi$PT~\cite{herrera97npb,kaiser00epjc}.
We will work in this combined expansion in our study and denote it as $\delta$ expansion throughout
the paper, where $\cO(\delta) \sim \cO(p^2) \sim \cO(m_q) \sim \cO(1/N_C)$.
This counting rule is different from the one proposed in Ref.~\cite{borasoy01epja}, where the $\eta_0$ mass is counted as $\cO(1)$ and the infrared
regularization method is employed to handle the chiral loops.

Some recent works in Refs.~\cite{herrera98plb,Gerard:2004gx,Gerard:2009ps,Mathieu:2010ss,Guo:2011pa} have addressed the $\eta$-$\eta'$ mixing
in the chiral framework up to next-to-leading order (NLO). As an improvement, we will perform the systematic study of the $\eta$-$\eta'$
mixing in the $\delta$-expansion scheme up to next-to-next-to-leading order (NNLO) and take into account the very recent lattice simulation data,
which are not considered in the previous works~\cite{herrera98plb,Gerard:2004gx,Gerard:2009ps,Mathieu:2010ss,Guo:2011pa}.
In addition, we also simultaneously analyze the $m_\pi$ dependences of other physical quantities from lattice simulations,
such as the axial $\pi, K$ decay constants and the mass ratio of the strange and up/down quarks, in order to further constrain the $\chi$PT LECs.

This article is organized as follows. In Sect.~\ref{sec.theoframe}, we introduce the theoretical framework and calculate the relevant physical
quantities. In Sect.~\ref{sec.pheno}, the phenomenological discussions will be presented. Conclusions will be given in Sect.~\ref{sec.concl}.
Further details about the calculations up to NNLO are relegated to App.~\ref{app.delta}.

\section{Theoretical framework }\label{sec.theoframe}

\subsection{Relevant chiral Lagrangian}

At leading order in the $\delta$ expansion, i.e. $\cO(\delta^0)$, the $U(3)$ $\chi$PT Lagrangian consists of three operators
\begin{eqnarray}\label{laglo}
\mL^{(\delta^0)}= \frac{ F^2}{4}\bra u_\mu u^\mu \ket+
\frac{F^2}{4}\bra \chi_+ \ket
+ \frac{F^2}{12}M_0^2 X^2 \,,
\end{eqnarray}
where the  chiral building blocks are defined as~\cite{gasser8485,kaiser00epjc,herrera97npb,herrera98plb}
\begin{eqnarray}\label{defbb}
&& U =  u^2 = e^{i\frac{ \sqrt2\Phi}{ F}}\,, \qquad \chi = 2 B (s + i p) \,,\qquad \chi_\pm  = u^\dagger  \chi u^\dagger  \pm  u \chi^\dagger  u \,,
\qquad    X= \log{(\det U)}\,,  \nn\\
&& u_\mu = i u^\dagger  D_\mu U u^\dagger \,, \qquad  D_\mu U \, =\, \partial_\mu U - i (v_\mu + a_\mu) U\, + i U  (v_\mu - a_\mu) \,,
\end{eqnarray}
with the pNGB octet+singlet matrix
\begin{equation}\label{phi1}
\Phi \,=\, \left( \begin{array}{ccc}
\frac{1}{\sqrt{2}} \pi^0+\frac{1}{\sqrt{6}}\eta_8+\frac{1}{\sqrt{3}} \eta_0 & \pi^+ & K^+ \\ \pi^- &
\frac{-1}{\sqrt{2}} \pi^0+\frac{1}{\sqrt{6}}\eta_8+\frac{1}{\sqrt{3}} \eta_0   & K^0 \\  K^- & \overline{K}^0 &
\frac{-2}{\sqrt{6}}\eta_8+\frac{1}{\sqrt{3}} \eta_0
\end{array} \right)\,,
\end{equation}
and $s,p,v_\mu, a_\mu$ being the external
scalar, pseudoscalar, vector and axial-vector sources, respectively.
The coupling $F$  appearing in Eqs.\eqref{laglo} and \eqref{defbb} corresponds to the pNGB axial decay constant
in the large $N_C$ and chiral limits. The light quark masses are introduced
by setting $(s+ip)=$diag$\{ \hat{m}, \hat{m}, m_s\}$,
being $\hat{m}$ the averaged up and down quark masses and $m_s$ that of the strange quark.

Notice the structure of the LO Lagrangian in Eq.~(\ref{laglo}):
the first operator is of $\cO({N_C,p^2})$ type, the second one corresponds to the type of
$\cO({N_C,m_q})$ and the last one  stems from  the QCD $U(1)_A$ anomaly and is of $\cO({N_C^0,p^0 })$ type,
where $U$ is counted as $\cO(1)$,$F^2\sim \cO(N_C)$ and $M_0^2\sim \cO(N_C^{-1})$
in the classification $\cO(N_C^j, p^k,m_q^\ell)$ of the EFT Lagrangian operators in Eq.~(\ref{laglo}).
In the following, we will denote the chiral expansions in powers of squared momenta $p^2$ and quark masses $m_q$ simply as a generic
expansion in $p^2$.
%%%, as done in most of the chiral study.

The NLO $U(3)$ chiral Lagrangian, i.e., $\cO(\delta)$,
contains $\cO({N_C,p^4})$ and $\cO({N_C^0,p^2})$ operators.
The relevant ones in our work read~\cite{kaiser00epjc}
\begin{eqnarray}\label{lagnlo}
\mL^{(\delta)} =&&  L_5 \bra  u^\mu u_\mu \chi_+ \ket
+\frac{ L_8}{2} \bra  \chi_+\chi_+ + \chi_-\chi_- \ket
+\frac{F^2\, \Lambda_1}{12}   D^\mu X D_\mu X  -\frac{F^2\, \Lambda_2}{12} X \bra \chi_- \ket\, ,
\end{eqnarray}
with the dimensionless LECs' scaling like $L_5, L_8 \sim \cO(N_C)$ and
$\Lambda_1,\Lambda_2 \sim \cO(N_C^{-1})$.

At NNLO, i.e. $\cO(\delta^2)$, there are three types of operators: $\cO(N_C^{-1},p^2)$,
$\cO(N_C^{0},p^4)$ and $\cO(N_C,p^6)$. Their explicit forms read~\cite{herrera97npb,bijnens99jhep}
\begin{eqnarray}\label{lagnnlo}
\mL^{(\delta^2)} =&& \frac{F^2\, v_2^{(2)}}{4} X^2 \bra \chi_+ \ket
\nn \\ &&
+  L_4 \bra  u^\mu u_\mu \ket \bra \chi_+ \ket
+ L_6 \bra  \chi_+ \ket  \bra  \chi_+ \ket
+ L_7 \bra  \chi_- \ket  \bra  \chi_- \ket
+  L_{18} \bra u_\mu \ket \bra u^\mu \chi_+ \ket
+  L_{25} X \bra \chi_+ \chi_- \ket
\nn \\ &&
+ C_{12} \bra   h_{\mu\nu}h^{\mu\nu}  \chi_+   \ket
+ C_{14} \bra u_\mu  u^\mu \chi_+ \chi_+  \ket
+ C_{17} \bra u_\mu  \chi_+ u^\mu \chi_+  \ket
\nn \\ &&
+ C_{19} \bra \chi_+ \chi_+ \chi_+  \ket
+ C_{31} \bra  \chi_-  \chi_- \chi_+ \ket\,,
\end{eqnarray}
where the first line corresponds to the $\cO(N_C^{-1},p^2)$ type, the second line is of the
$\cO(N_C^{0},p^4)$ type and the last two lines are of the $\cO(N_C,p^6)$ type.
The LECs carry the scalings  $v_2^{(2)}\sim \cO(N_C^{-2})$,
$L_4,L_6, L_7, L_{18}, L_{25}\sim \cO(N_C^0)$
and $C_{12}, C_{14}, C_{17}, C_{19}, C_{31}\sim \cO(N_C)$.
Notice that we have only shown the
operators at different $\delta$ orders in Eqs.~\eqref{laglo}, \eqref{lagnlo} and \eqref{lagnnlo} that are pertinent
to our present study, not aiming at giving the complete sets of operators.
The conventions to label the LO, NLO and NNLO operators in Eqs.~\eqref{laglo},~\eqref{lagnlo} and \eqref{lagnnlo} follow closely
the notations in Refs.~\cite{herrera97npb,kaiser00epjc,bijnens99jhep}.
Unless it is explicitly stated, the LECs will correspond to $U(3)$ $\chi$PT and must not be confused
with those in $SU(3)$ $\chi$PT. The matching between these two EFTs can be found in Ref.~\cite{kaiser00epjc}.
 The terms $L_{j}$ are denoted as $\beta_j$ in Ref.~\cite{herrera97npb,herrera98plb}.

Comparing the $U(3)$ and $SU(3)$ theories one can observe that
some terms have been reshuffled in the $\delta$ expansion of the $U(3)$ Lagrangian.
For example, the $ L_{i=4,5,6,7,8}$ terms are NLO in $SU(3)$ $\chi$PT,
but they are now split into NLO and NNLO in the $\delta$ expansion  (see Eqs.~\eqref{lagnlo} and \eqref{lagnnlo}).
We have several additional new operators, namely the last one in Eq.~\eqref{laglo}, the $\Lambda_{i=1,2}$ in Eq.~\eqref{lagnlo} and
the $v_2^{(2)},  L_{18},  L_{25}$ terms in Eq.~\eqref{lagnnlo}, that are absent in the $SU(3)$ $\chi$PT case.
Finally, the chiral loops start contributing at NNLO in the $\delta$ expansion, while they appear at NLO
in the conventional $SU(3)$ case.

\subsection{ The $\eta$-$\eta'$ mixing at NNLO in $\delta$ expansion}

Next we calculate the $\eta$-$\eta'$ mixing order by order in the $\delta$ expansion. In literature, there are two
bases to address the $\eta$-$\eta'$ mixing, namely the singlet-octet basis with $\eta_0$ and $\eta_8$, and the
quark-flavor basis with $\eta_q$ and $\eta_s$. The relations between fields in these two bases are
\begin{equation}\label{etaqstoeta08}
\left(\begin{array}{c} \eta_8 \\ \eta_0 \end{array}\right) \quad=\quad
\left(\begin{array}{cc} \sqrt{\frac{1}{3}} & - \sqrt{\frac{2}{3}} \\  \sqrt{\frac{2}{3}} & \sqrt{\frac{1}{3}} \end{array}\right) \quad
\left(\begin{array}{c} \eta_q \\ \eta_s\end{array}\right) \, .
%%%
%%%\eta_q=\sqrt{\frac{1}{3}}\eta_8+ \sqrt{\frac{2}{3}} \eta_0\,, \qquad \eta_s=-\sqrt{\frac{2}{3}}\eta_8+ \sqrt{\frac{1}{3}} \eta_0\,.
%%%
\end{equation}
In the large--$N_C$ limit where the $U(1)_A$ anomaly is absent,  $\eta_q$ and $\eta_s$
are the mass eigenstates and they are generated by the axial-vector currents with the
quark flavors $q\bar{q}=(u\bar{u}+d\bar{d})/\sqrt2$ and $s\bar{s}$, respectively.
The two bases are related to each other through an orthogonal transformation
and provide an equivalent description for the $\eta$-$\eta'$ mixing.

%%%
%%%We start the discussion with the $\eta_0$-$\eta_8$ basis.
%%%
As noticed in Refs.~\cite{Guo:2011pa,Guo:2012yt},
when doing the loop calculations with $\eta$ and $\eta'$, it
is rather cumbersome to work with the $\eta_0$ and $\eta_8$ states. The reason is that at leading order
the Lagrangian in Eq.~\eqref{laglo} gives the mixing between $\eta_0$ and $\eta_8$, and the mixing strength is proportional to
$m_K^2-m_\pi^2$, which in the $\delta$ expansion is formally counted as the same order as the
diagonal terms in the mass matrix for $\eta_0$ and $\eta_8$.
As a result, the insertion of the $\eta_0$-$\eta_8$ mixing in the chiral loops will not increase the $\delta$ order of the loop diagrams.
This  makes the loop calculation technically much more complicated, as one needs to consider the arbitrary insertions
of the $\eta_0$-$\eta_8$ mixing in the chiral loop diagrams. Nevertheless, Refs.~\cite{Guo:2011pa,Guo:2012yt} provide a simple recipe
to handle this problem by expressing the Lagrangian in terms of the $\overline{\eta}$ and $\overline{\eta}'$ states
which result from the diagonalization of $\eta_0$ and $\eta_8$ at leading order in $\delta$.
The main difference is that
the mixing between $\overline{\eta}$ and $\overline{\eta}'$ is now at least a NLO effect in $\delta$,
while the $\eta_0$-$\eta_8$ mixing was appearing at LO. The relation between
the LO mass eigenstates $\bar{\eta}$ and $\bar{\eta}'$ and the singlet-octet basis is given by
the mixing angle $\theta$:
\begin{eqnarray}\label{lomixing}
\left(\begin{array}{c} \bar\eta \\ \bar{\eta}' \end{array}\right) \quad=\quad
\left(\begin{array}{cc} c_\theta & - s_\theta \\  s_\theta & c_\theta \end{array}\right) \quad
\left(\begin{array}{c} \eta_8 \\ \eta_0\end{array}\right) \, ,
%%%
%%%\overline{\eta}=c_\theta \eta_8 - s_\theta\eta_0\,,  \quad
%%%\overline{\eta}'= s_\theta \eta_8 + c_\theta  \eta_0\,,
%%%
\end{eqnarray}
with $c_\theta=\cos{\theta}$ and $s_\theta=\sin{\theta}$. The LO mixing angle $\theta$
and masses of $\overline{\eta}$ and $\overline{\eta}'$
 are given by the leading order Lagrangian $\mL^{(\delta^0)}$ in Eq.~\eqref{laglo} (see e.g. Ref.~\cite{Guo:2011pa}):
\begin{eqnarray}
m_{\overline{\eta}}^2 &=& \frac{M_0^2}{2} + \overline{m}_K^2
- \frac{\sqrt{M_0^4 - \frac{4 M_0^2 \Delta^2}{3}+ 4 \Delta^4 }}{2} \,, \label{defmetab2}  \\
m_{\overline{\eta}'}^2 &=& \frac{M_0^2}{2} + \overline{m}_K^2
+ \frac{\sqrt{M_0^4 - \frac{4 M_0^2 \Delta^2}{3}+ 4 \Delta^4 }}{2} \,, \label{defmetaPb2}  \\
\sin{\theta} &=& -\left( \sqrt{1 +
\frac{ \big(3M_0^2 - 2\Delta^2 +\sqrt{9M_0^4-12 M_0^2 \Delta^2 +36 \Delta^4 } \big)^2}{32 \Delta^4} } ~\right )^{-1}\,,
\label{deftheta0}
\end{eqnarray}
with $\Delta^2 = \overline{m}_K^2 - \overline{m}_\pi^2$. Here $\overline{m}_K$ and $\overline{m}_\pi$ denote
the LO  kaon and pion masses, respectively.

When higher order corrections are taken into account, the LO diagonalized $\overline{\eta}$ and $\overline{\eta}'$ will get mixed
again. Up to the NNLO,  a general parametrization of the bilinear terms involving the $\overline{\eta}$ and $\overline{\eta}'$
states can be written as
\begin{eqnarray}\label{lagmixingpara}
\mL&=&   \frac{\delta_1}{2} \,\partial_\mu\partial_\nu \overline{\eta} \partial^\mu\partial^\nu \overline{\eta}
+\frac{\delta_2}{2} \,\partial_\mu\partial_\nu \overline{\eta}' \partial^\mu\partial^\nu \overline{\eta}'
+\delta_3 \,\partial_\mu\partial_\nu \overline{\eta} \partial^\mu\partial^\nu \overline{\eta}'  \nn \\ & &
+\frac{1 + \delta_{\overline{\eta}} }{2}\partial_\mu \overline{\eta} \partial^\mu\overline{\eta}
+\frac{1+ \delta_{\overline{\eta}'} }{2}\partial_\mu \overline{\eta}' \partial^\mu \overline{\eta}'
+ \delta_k\, \partial_\mu \overline{\eta} \partial^\mu \overline{\eta}'
\nonumber \\ & &
-\frac{m_{\overline{\eta}}^2 + \delta_{m_{\overline{\eta}}^2} }{2} \overline{\eta}\, \overline{\eta}
- \frac{m_{\overline{\eta}'}^2 + \delta_{m_{\overline{\eta}'}^2 } }{2} \overline{\eta}' \overline{\eta}'
- \delta_{m^2} \,\overline{\eta}\, \overline{\eta}' \,,
\end{eqnarray}
where the $\delta_i's$ contain the NLO and NNLO corrections.
Here these operators must be understood as the terms of the effective action that provide
the pseudoscalar meson self-energies.
The higher-derivative terms $\delta_{j=1,2,3}$ in the first
line of Eq.~\eqref{lagmixingpara} are exclusively contributed by the $\cO(p^6)$ operator $C_{12}$ in Eq.~\eqref{lagnnlo},
which belongs to the NNLO Lagrangian. The remaining $\delta_i's$
receive contributions from the NLO operators in Eq.~\eqref{lagnlo}, the NNLO ones in Eq.~\eqref{lagnnlo} and the one-loop diagrams,
which contribute at NNLO.   Their explicit expressions can be found in App.~\ref{app.delta}.

At leading order, there is only the mass mixing term from Eq.~\eqref{laglo} whereas at NLO and NNLO
one has to deal in addition with the kinematic mixing terms in Eq.~\eqref{lagmixingpara}, apart from the mass mixing. The physical states of $\eta$ and $\eta'$
can be obtained from the perturbative-expansion ($\delta$-expansion) in three steps:
as a first step, we eliminate the higher-derivative terms through the field
redefinitions of $\overline{\eta}$ and $\overline{\eta}'$;
then we transform and rescale the fields resulting from the first step in order to write the kinematic terms in the canonical form;
after the preceding two steps, there is only the mass mixing term left, which is straightforward to handle.

In the first step, we make the following field redefinitions for the $\overline{\eta}$ and $\overline{\eta}'$ states
\begin{eqnarray}\label{deletehd}
\overline{\eta}  \rightarrow \overline{\eta} + \alpha_1\square \overline{\eta}+ \alpha_2\square \overline{\eta}'\,,  &&
\overline{\eta}' \rightarrow \overline{\eta}' + \alpha_2\square \overline{\eta}+ \alpha_3\square \overline{\eta}'\,,
\end{eqnarray}
with the d'Alembert operator $\square\equiv\partial_\mu\partial^\mu$.
After some algebra manipulations, it is straightforward to obtain
\begin{equation}
\alpha_1 = -\frac{\delta_1}{2} \,,  \qquad \alpha_2 = -\frac{\delta_3}{2} \,,  \qquad \alpha_3 = -\frac{\delta_2}{2} \,,
\end{equation}
so that the three higher-derivative terms in Eq.~\eqref{lagmixingpara} will be eliminated.
Notice that the $\alpha_{1,2,3}$ are NNLO, i.e., $\cO(\delta^2)$.
Substituting the field redefinitions from Eq.~\eqref{deletehd} into the general mixing structure in Eq.~\eqref{lagmixingpara} and
keeping the terms up to NNLO, the resulting bilinear Lagrangian reads
\begin{eqnarray}\label{lagmixingpara2}
\mL&=&    \frac{1 + \delta_{\overline{\eta}} + m_{\overline{\eta}}^2\delta_1  }{2}\partial_\mu \overline{\eta} \partial^\mu\overline{\eta}
+\frac{1+ \delta_{\overline{\eta}'} + m_{\overline{\eta}'}^2\,\delta_2 }{2}\partial_\mu \overline{\eta}' \partial^\mu \overline{\eta}'
+ \big[\delta_k+\frac{\delta_3}{2}(m_{\overline{\eta}}^2+m_{\overline{\eta}'}^2) \big]\, \partial_\mu \overline{\eta} \partial^\mu \overline{\eta}'
\nonumber \\ & &
-\frac{m_{\overline{\eta}}^2 + \delta_{m_{\overline{\eta}}^2} }{2} \overline{\eta}\, \overline{\eta}
- \frac{m_{\overline{\eta}'}^2 + \delta_{m_{\overline{\eta}'}^2 } }{2} \overline{\eta}' \overline{\eta}'
- \delta_{m^2} \,\overline{\eta}\, \overline{\eta}' \,.
\end{eqnarray}

In the second step, we need to eliminate the kinematic mixing term in Eq.~\eqref{lagmixingpara2}, and then to rescale the
fields to have them in the canonical forms. This can be done perturbatively. In the final step, we take care of the mass mixing term.
The last two steps can be achieved through the following field transformations
\begin{equation}\label{transnnlo}
\left(
\begin{array}{c}
\eta \\ \eta'
\end{array}
\right)=\left(\begin{array}{cc} \cos\theta_{\delta} & -\sin\theta_{\delta}\\
\sin\theta_{\delta} & \cos\theta_{\delta}
\end{array}\right)
\left(\begin{array}{cc}
1+\delta_A & \delta_B
\\
\delta_B &
1+\delta_C
\end{array}\right)
\left(\begin{array}{c}
\overline{\eta} \\ \overline{\eta}'
\end{array}\right)\,,
\end{equation}
with $\eta, \eta'$ the physical states and
\begin{eqnarray}
 \delta_{A} &=& \frac{\delta_{\overline{\eta}}}{2}+ \frac{m_{\overline{\eta}}^2\,\delta_1}{2}-\frac{\delta_{\overline{\eta},{\rm NLO}}^{2}}{8}
-\frac{\delta_{k,{\rm NLO}}^{2}}{8} \,, \nn\\
 \delta_{B} &=& \frac{\delta_{k}}{2}+\frac{\delta_3}{4}(m_{\overline{\eta}}^2+m_{\overline{\eta}'}^2)
 -\frac{\delta_{\overline{\eta},{\rm NLO}}\delta_{k,{\rm NLO}}}{8}-\frac{\delta_{\overline{\eta}',{\rm NLO}}\delta_{k,{\rm NLO}}}{8}
 \,, \nn \\
 \delta_{C} &=& \frac{\delta_{\overline{\eta}'}}{2}+ \frac{ m_{\overline{\eta}'}^2 \delta_2 }{2}
 -\frac{\delta_{\overline{\eta}',{\rm NLO}}^{2}}{8}-\frac{\delta_{k,{\rm NLO}}^{2}}{8}\,,
\end{eqnarray}
where $\delta_{\overline{\eta},{\rm NLO}}, \delta_{\overline{\eta}',{\rm NLO}}, \delta_{k,{\rm NLO}}$
stand for the NLO parts of the three quantities respectively.
We point out that $\delta_{\overline{\eta}}, \delta_{\overline{\eta}'}, \delta_{k}$ receive both NLO and NNLO contributions, while $\delta_1,\delta_2,\delta_3$
are only contributed by the NNLO effect, which is the $C_{12}$ operator in Eq.~\eqref{lagnnlo}.
Comparing with the NLO results in Eq.~(15) from our previous paper~\cite{Guo:2011pa}, we have generalized the
expression to the NNLO case in the present Eq.~\eqref{transnnlo}. 
Another way to treat the mixing of pseudoscalar mesons 
in $\chi$PT was also previously studied in Ref.~\cite{Amoros:2001cp} and applied to the $\pi^0$-$\eta$ case up to the two-loop level.

In the practical calculation, it is more often to use the inverse of the relations in Eq.~\eqref{transnnlo},
where the perturbative expansion leads to
\begin{equation}\label{transnnlo2}
\left(\begin{array}{c}
\overline{\eta} \\ \overline{\eta}'
\end{array}\right)
=
\left(\begin{array}{cc}
1+\delta_A' & \delta_B'
\\
\delta_B' &
1+\delta_C'
\end{array}\right)
\left(\begin{array}{cc} \cos\theta_{\delta} & \sin\theta_{\delta}\\
-\sin\theta_{\delta} & \cos\theta_{\delta}
\end{array}\right)
\left(
\begin{array}{c}
\eta \\ \eta'
\end{array}
\right)\,,
\end{equation}
with
\begin{eqnarray}
 \delta_{A}' &=& -\frac{\delta_{\overline{\eta}}}{2}- \frac{m_{\overline{\eta}}^2\,\delta_1}{2} + \frac{3\delta_{\overline{\eta},{\rm NLO}}^{2}}{8}
+\frac{3\delta_{k,{\rm NLO}}^{2}}{8} \,, \nn\\
 \delta_{B}' &=& -\frac{\delta_{k}}{2}-\frac{\delta_3}{4}(m_{\overline{\eta}}^2+m_{\overline{\eta}'}^2)
 +\frac{3\delta_{\overline{\eta},{\rm NLO}}\delta_{k,{\rm NLO}}}{8} +\frac{3\delta_{\overline{\eta}',{\rm NLO}}\delta_{k,{\rm NLO}}}{8} \,, \nn \\
 \delta_{C}' &=& -\frac{\delta_{\overline{\eta}'}}{2} - \frac{ m_{\overline{\eta}'}^2 \delta_2 }{2}
 +\frac{3\delta_{\overline{\eta}',{\rm NLO}}^{2}}{8}+\frac{3\delta_{k,{\rm NLO}}^{2}}{8}\,.
\end{eqnarray}

The $\theta_{\delta}$ appearing in Eqs.~\eqref{transnnlo} and \eqref{transnnlo2} is determined through
\begin{equation}
\tan\theta_{\delta}=\frac{ \widehat{\delta}_{m^{2}}}{m_{\eta'}^{2}-\widehat{m}_{\eta}^{2}}\,,
\end{equation}
with
\begin{eqnarray}
\widehat{\delta}_{m^{2}} &=&
\delta_{m^{2}}
-\frac{1}{2}\left[\delta_{k}+\frac{\delta_3}{2}(m_{\overline{\eta}}^2+m_{\overline{\eta}'}^2)  \right]
\left( m_{\overline{\eta}}^{2}+m_{\overline{\eta}'}^{2}\right)
+\frac{1}{8}\delta_{k,{\rm NLO}}\delta_{\overline{\eta},{\rm NLO}}\left( 5m_{\overline{\eta}}^{2}+3m_{\overline{\eta}'}^{2}\right)
\nn \\  &&
-\frac{1}{2}\delta_{k,{\rm NLO}}\left( \delta_{m_{\overline{\eta}}^{2},{\rm NLO}}+\delta_{m_{\overline{\eta}'}^{2},{\rm NLO}}\right)
+\frac{1}{8}\delta_{k,{\rm NLO}}\delta_{\overline{\eta}',{\rm NLO}}\left( 3m_{\overline{\eta}}^{2}+5m_{\overline{\eta}'}^{2}\right)
\nn \\  &&
-\frac{1}{2}\delta_{m^{2},{\rm NLO}}\left( \delta_{\overline{\eta},{\rm NLO}}+\delta_{\overline{\eta}',{\rm NLO}}\right) \,,
\nn \\
\widehat{m}_{\eta}^{2} &=& m_{\overline{\eta}}^{2}+\delta_{m_{\overline{\eta}}^{2}}
-m_{\overline{\eta}}^{2}\left(\delta_{\overline{\eta}} + m_{\overline{\eta}}^2\delta_1  \right)+m_{\overline{\eta}}^{2}\delta_{\overline{\eta},{\rm NLO}}^{2}
+\frac{3}{4}m_{\overline{\eta}}^{2}\delta_{k,{\rm NLO}}^{2}+\frac{1}{4}m_{\overline{\eta}'}^{2}\delta_{k,{\rm NLO}}^{2}
 \nn \\&&
-\delta_{k,{\rm NLO}}\delta_{m^{2},{\rm NLO}}-\delta_{\overline{\eta},{\rm NLO}}\delta_{m_{\overline{\eta}}^{2},{\rm NLO}} \,,\nn \\
\widehat{m}_{\eta'}^{2} &=& m_{\overline{\eta}'}^{2}+\delta_{m_{\overline{\eta}'}^{2}}
-m_{\overline{\eta}'}^{2}\left(\delta_{\overline{\eta}'}+ m_{\overline{\eta}'}^2\delta_2 \right)+m_{\overline{\eta}'}^{2}
\delta_{\overline{\eta}',{\rm NLO}}^{2}
+\frac{1}{4}m_{\overline{\eta}}^{2}\delta_{k,{\rm NLO}}^{2}+\frac{3}{4}m_{\overline{\eta}'}^{2}\delta_{k,{\rm NLO}}^{2}
 \nn \\ &&
-\delta_{k,{\rm NLO}}\delta_{m^{2},{\rm NLO}}-\delta_{\overline{\eta}',{\rm NLO}}\delta_{m_{\overline{\eta}'}^{2},{\rm NLO}} \,, \nn \\
2m_{\eta}^{2} &=& \widehat{m}_{\eta}^{2}+\widehat{m}_{\eta'}^{2}-\sqrt{\left( \widehat{m}_{\eta}^{2}
-\widehat{m}_{\eta'}^{2}\right)^{2}+4\widehat{\delta}_{m^2}^{2} } \,,\nn \\
2m_{\eta'}^{2} &=& \widehat{m}_{\eta}^{2}+\widehat{m}_{\eta'}^{2}+\sqrt{\left( \widehat{m}_{\eta}^{2}-\widehat{m}_{\eta'}^{2}\right)^{2}
+4 \widehat{\delta}_{m^{2}}^{2} }\,,
\end{eqnarray}
where $\delta_{i,{\rm NLO}}$ stand for the NLO parts of $\delta_{i}$.

In the phenomenological discussions, the popular two-mixing-angle parametrization in the singlet-octet
basis~\cite{leutwyler98npbsp,Kaiser:1998ds} takes the form
\begin{eqnarray} \label{twoanglesmixing08}
 \left(
 \begin{array}{c}
 \eta   \\
 \eta' \\
 \end{array}
 \right) = \frac{1}{F}\left(
                                        \begin{array}{cc}
                                          F_8\, \cos{\theta_8}  & -F_0 \,\sin{\theta_0}  \\
                                           F_8\,\sin{\theta_8} & F_0 \,\cos{\theta_0} \\
                                        \end{array}
    \right)
      \left(
       \begin{array}{c}
       \eta_8   \\
       \eta_0  \\
       \end{array}
        \right)\,.
\end{eqnarray}

Combining Eqs.~\eqref{lomixing} and~\eqref{transnnlo},
it is straightforward to derive the relations between the four parameters in the two-mixing-angle scheme
in Eq.~\eqref{twoanglesmixing08} and the $\chi$PT LECs:
\begin{eqnarray}
 F_8^2 = && F^2 \bigg\{ \big[\cos(\theta+\theta_{\delta}) + \delta_B \sin(\theta-\theta_{\delta})
+ \delta_A \cos\theta \cos\theta_\delta - \delta_C \sin\theta \sin\theta_\delta \big]^2
\nn \\ &&  \qquad
+ \big[ \sin(\theta+\theta_{\delta}) + \delta_B \cos(\theta-\theta_{\delta})
+ \delta_A \cos\theta \sin\theta_\delta + \delta_C \sin\theta \cos\theta_\delta  \big]^2
\bigg\} \,,
\nn \\
 F_0^2 = && F^2 \bigg\{ \big[ -\sin(\theta+\theta_{\delta}) + \delta_B \cos(\theta-\theta_{\delta})
- \delta_A \sin\theta \cos\theta_\delta - \delta_C \cos\theta \sin\theta_\delta  \big]^2
\nn \\ &&  \qquad
+ \big[ \cos(\theta+\theta_{\delta}) - \delta_B \sin(\theta-\theta_{\delta})
- \delta_A \sin\theta \sin\theta_\delta + \delta_C \cos\theta \cos\theta_\delta  \big]^2
\bigg\} \,,
\nn \\
 \tan\theta_8 = &&  \dfrac{ \sin(\theta+\theta_{\delta}) + \delta_B \cos(\theta-\theta_{\delta})
+ \delta_A \cos\theta \sin\theta_\delta + \delta_C \sin\theta \cos\theta_\delta }
{ \cos(\theta+\theta_{\delta}) + \delta_B \sin(\theta-\theta_{\delta})
+ \delta_A \cos\theta \cos\theta_\delta - \delta_C \sin\theta \sin\theta_\delta} \,,
\nn \\
 \tan\theta_0 = && - \dfrac{-\sin(\theta+\theta_{\delta}) + \delta_B \cos(\theta-\theta_{\delta})
- \delta_A \sin\theta \cos\theta_\delta - \delta_C \cos\theta \sin\theta_\delta }
{ \cos(\theta+\theta_{\delta}) - \delta_B \sin(\theta-\theta_{\delta})
- \delta_A \sin\theta \sin\theta_\delta + \delta_C \cos\theta \cos\theta_\delta }\,,
\end{eqnarray}
where the $\chi$PT LECs are implicitly included in $\theta, \theta_\delta$, $\delta_A$, $\delta_B$ and $\delta_C$.
Since $\theta_\delta,\delta_A,\delta_B,\delta_C\sim \cO(\delta)$ or $\cO(\delta^2)$, at LO one has $F_8=F_0=F$
and one mixing-angle $\theta_8=\theta_0=\theta$.

The relations between the physical $\eta, \eta'$ states and the quark-flavor basis is commonly
parametrized as
\begin{eqnarray} \label{twoanglesmixingqs}
 \left(
 \begin{array}{c}
 \eta   \\
 \eta' \\
 \end{array}
 \right) = \frac{1}{F}\left(
                                        \begin{array}{cc}
                                          F_q\, \cos{\phi_q}  & -F_s \,\sin{\phi_s}  \\
                                           F_q\,\sin{\phi_q} & F_s \,\cos{\phi_s} \\
                                        \end{array}
    \right)
      \left(
       \begin{array}{c}
       \eta_q   \\
       \eta_s  \\
       \end{array}
        \right)\,.
\end{eqnarray}
Combining Eqs.~\eqref{etaqstoeta08},~\eqref{lomixing} and~\eqref{transnnlo},
%%%~\eqref{twoanglesmixing08} and ,
it is straightforward to obtain the parameters in Eq.~\eqref{twoanglesmixingqs}:
\begin{eqnarray}\label{fqfstof0f8}
F_q^2 &=& \frac{2F_0^2 + F_8^2 -2\sqrt2 F_0 F_8 \sin(\theta_0-\theta_8)}{3}\,, \nn \\
F_s^2 &=& \frac{ F_0^2 + 2F_8^2 + 2\sqrt2 F_0 F_8 \sin(\theta_0-\theta_8)}{3}\,, \nn \\
\tan\phi_q &=& \frac{ \sqrt2 F_8 \cos\theta_8 + F_0 \sin\theta_0}{ \sqrt2 F_0 \sin\theta_0 - F_8 \cos\theta_8}\,, \nn \\
\tan\phi_s &=& \frac{ \sqrt2 F_0 \cos\theta_0 + F_8 \sin\theta_8}{ \sqrt2 F_8 \sin\theta_8 - F_0 \cos\theta_0}\,,
\end{eqnarray}
where at LO in the $\delta$--expansion one has $F_q=F_s=F$ and $\phi_q=\phi_s=\theta_{\rm id} - \theta$,
with the ideal mixing $\theta_{\rm id}= - \arcsin\sqrt{2/3}$.

\subsection{Insights into previous  studies of the  $\eta$-$\eta'$ mixing}

In the previous subsection we have performed the full computation of the mixing up to NNLO in the $\delta$ expansion.
It is interesting to make a brief summary of the assumptions made in previous works, where
plenty of mixing formalisms have been proposed to address
the $\eta$-$\eta'$
system~\cite{herrera98plb,Gerard:2004gx,Gerard:2009ps,Mathieu:2010ss, feldmann98prd,Georgi:1993jn,Peris:1993np,Guo:2011pa}.
In Ref.~\cite{Georgi:1993jn}, only the lowest order in the quark masses and $1/N_C$, i.e. the LO contributions in the $\delta$ expansion,
were taken into account.
 Even though it provided a reasonable first approximation, it failed to give an accurate description
of the experimentally observed mass ratio $m_\eta^2/m_{\eta'}^2$.
The $\cO(p^2)$ contributions were studied up to NLO in $1/N_C$ in Ref.~\cite{Peris:1993np}
(including the terms in Eq.~\eqref{laglo} and $\Lambda_1$ and $\Lambda_2$ in Eq.~\eqref{lagnlo}), perfectly explaining
the experimental value of $m_\eta^2/m_{\eta'}^2$.
However, it turned out to be inadequate to give a proper value for the $\eta$-$\eta'$ mixing angle.
On the other hand, the authors in Refs.~\cite{Gerard:2004gx,Gerard:2009ps,Mathieu:2010ss}
went up to NLO in the $p^2$ expansion but keeping just the LO in $1/N_C$
(including the terms in Eq.~\eqref{laglo} and $L_5$ and $L_8$ in Eq.~\eqref{lagnlo}).
Both the $\eta$-$\eta'$ mixing angle and the ratio $F_K/F_\pi$ were qualitatively reproduced in this case.
The full set of NLO contributions in the $\delta$--expansion (i.e., the effects up to NLO both in $1/N_C$ and $p^2$)
was analyzed in Ref.~\cite{herrera98plb}, together with the mixing angle and the $\pi, K, \eta$ and $\eta'$ axial-vector decay constants.
In Ref.~\cite{Guo:2011pa}, the contributions from the tree-level resonance exchanges and partial NNLO effects, e.g. the loop diagrams,
were considered for the masses of $\eta$ and $\eta'$. In this work, we generalize the discussions up to the full
NNLO study in the $\delta$--expansion and confront our theoretical expressions with the very
recent lattice simulation data and the phenomenological inputs from the two-mixing-angle scheme.

Reference~\cite{feldmann98prd} introduced a quark-model inspired approach to the $\eta$-$\eta'$ mixing, which
is commonly referred as the FKS formalism and
used in many phenomenological analyses~\cite{Feldmann:1999uf}. The essence of the FKS formalism is the assumption that the
axial decay constants in the quark-flavor basis takes the same mixing pattern as the states
\begin{eqnarray}\label{fksform}
    \left(
                                        \begin{array}{cc}
                                          F^q_\eta   & F^s_\eta   \\
                                           F^q_{\eta'}   & F^s_{\eta'}   \\
                                        \end{array}
    \right) =
 \left(
                                        \begin{array}{cc}
                                          \cos\phi   &  -\sin\phi  \\
                                           \sin\phi & \cos\phi \\
                                        \end{array}
    \right)
    \left(
                                        \begin{array}{cc}
                                          F_q   & 0   \\
                                           0 & F_s   \\
                                        \end{array}
    \right)\,,
\end{eqnarray}
where the decay constants are defined as the matrix elements of the axial currents
\begin{eqnarray}
 \bra 0 | A^a_\mu(0) | P(k)\ket = i \sqrt2 F^a_P k_\mu\,, \qquad (a=q,s; P=\eta,\eta')\,, \nn \\
 A^q_\mu= \frac{1}{\sqrt2} \big( \bar{u} \gamma_\mu \gamma_5 u +  \bar{d} \gamma_\mu \gamma_5 d \big)\,,
 \qquad  A^s_\mu=  \bar{s} \gamma_\mu \gamma_5 s \,.
\end{eqnarray}

From another point of view, the pattern of Eq.~\eqref{fksform} employed in the FKS formalism relies on the assumption that
there is no mixing between the decay constants of the flavor states $\eta_q$ and $\eta_s$.
In the $\chi$PT framework, the physical masses and decay constants can be obtained from the bilinear parts of nonet fields in the effective action with
the correlation  function of two axial currents.  Since the correlation  function is the second derivative with respect to the axial-vector external source $a_\mu$, 
and $a_\mu$ always appears in the Lagrangian together with the partial derivative $\partial_\mu$ as shown in Eq.~\eqref{defbb},
the absence of the mixing for the $\eta_q$ and $\eta_s$ decay constants in Eq.~\eqref{fksform} implies
that there are no kinematic mixing terms for the quark-flavor states $\eta_q$ and $\eta_s$ in the FKS formalism.
In fact, the assumption in Ref.~\cite{Gerard:2004gx} is in accord with the FKS formalism. This can be simply demonstrated
by expanding the chiral operators considered in Refs.~\cite{Gerard:2004gx,Gerard:2009ps}, i.e. those in Eq.~\eqref{laglo} and
$ L_5,  L_8$ in Eq.~\eqref{lagnlo}, up to quadratic terms in $\eta_q$ and $\eta_s$. \footnote{Our $ L_5$ and $ L_8$ operators correspond to the $\Lambda_2$ and $\Lambda_1$ terms
in Refs.~\cite{Gerard:2004gx,Gerard:2009ps,Mathieu:2010ss}, respectively. The $\Lambda$ term in the previous references corresponds
to our $\Lambda_2$ operator in Eq.~\eqref{lagnlo}. The $\Lambda$ term, though introduced from the beginning in these references, is dropped
in their later discussions, since it is $1/N_C$ suppressed.}  
No kinematic mixing terms for the
$\eta_q$ and $\eta_s$ fields result from these chiral operators. This also confirms the finding in Ref.~\cite{Mathieu:2010ss} that only when the
NLO of $1/N_C$ operator is excluded the FKS formalism is recovered with their chiral Lagrangian calculations.

Since general terms up to NNLO in $\delta$ expansion are kept in our discussion, unlike in the previous
works~\cite{Georgi:1993jn,Peris:1993np,herrera98plb,Gerard:2004gx,Gerard:2009ps,Mathieu:2010ss,Guo:2011pa}
where different assumptions, such as the preference of the higher order $p^2$ and $1/N_C$ effects, are made,
it is important and interesting for us to justify these assumptions in later discussions.

\subsection{Masses and decay constants of pion and kaon up to NNLO in $\delta$ expansion} \label{sect.nnloformula}

The NLO expression of the pion decay constant in the $\delta$ expansion reads
 \begin{eqnarray} \label{fpinlof0}
 F_\pi = F \bigg( 1 + 4  L_5\frac{ m_\pi^2}{ F^2} \bigg) \, ,
\end{eqnarray}
or, up to the precision considered, one can also use the physical $F_\pi$ in the expression inside brackets,
\begin{eqnarray} \label{fpinlofpi}
 F_\pi = F \bigg( 1 + 4  L_5\frac{ m_\pi^2}{ F_\pi^2} \bigg) \, .
\end{eqnarray}
The differences between Eqs.~\eqref{fpinlof0} and~\eqref{fpinlofpi}  are NNLO effects.
We mention that at a given order there is always ambiguity in
choosing the renormalized quantities in the higher order expressions.
In contrast, there is formally no ambiguity in the expressions in terms of the quantity $F$, which is
the pNGB axial decay constant in the chiral and large $N_C$ limits.
For example, if we limit our analysis up to NLO, formally, it is equally good to use $F_\pi$ or $F_K$
in the denominators of the NLO part in Eq.~\eqref{fpinlofpi},
since the difference is beyond the NLO precision. A typical solution in the chiral study is to express the quantities,
such as $m_\pi, F_\pi, m_K, F_K$,
in terms of the renormalized $F_\pi$ in the higher order corrections, as done in
the two-loop calculations in $SU(3)$ $\chi$PT~\cite{Amoros:1999dp}.
We follow this rule throughout the current work to estimate the uncertainty due to the truncation of the $\delta$ expansion
when one works at a given order in perturbation theory.
We mention that the notation of $m_\pi^2$ in the above equations 
stands for the renormalized pion mass squared and the leading order mass squared is denoted by $\overline{m}_\pi^2$. 
Notice the LO pion mass squared $\overline{m}_\pi^2$ is the one that is linear in the quark masses. 
The expressions relating $m_\pi^2$ and $\overline{m}_\pi^2$ will be discussed below.

Similarly up to NNLO, we can either use $F$ or $F_\pi$ in the NLO and NNLO expressions for other quantities
such as $F_K$ and the $\delta_i$'s in Eq.~\eqref{lagmixingpara}.
In the NNLO expressions, the difference between using $F$ or $F_\pi$ in the denominators is
a next-to-next-to-next-to-leading order effect (N$^3$LO).
Since in this work we study lattice simulation data up to pion mass of 500~MeV,
the convergence of the chiral series is expected to be much slower than that in the physical case with $m_\pi=135$~MeV.
Therefore it is {\it a priori} not trivial  to judge whether the two approaches--using $1/F^2$ and $1/F_\pi^2$--
are numerically equivalent or the lattice data prefer one of them.
Indeed in Ref.~\cite{Bernard:2009ds}, it is already noticed that to use $F$ or $F_\pi$ could cause some noticeable effects.
We will use the difference between both approaches as an estimate of the truncation error
at a given order in $\delta$.

We take the pion decay constant as an example to illustrate the differences of using $F$ and $F_\pi$ in the higher order expressions.
Using  $F$ in the higher order corrections, its expression reads
\begin{eqnarray}\label{fpinnlof0}
 F_\pi =&& F \bigg[ 1 + 4  L_5\frac{ m_\pi^2}{ F^2}
 + 4  L_4 \frac{ m_\pi^2 + 2m_K^2 }{ F^2}  +
 (24 L_5^2 - 64  L_5 L_8)\frac{m_\pi^4}{F^4}
 +  (8 C_{14} + 8 C_{17} )\frac{m_\pi^4}{F^2}
\nn \\ && \qquad
 + \frac{A_0(m_\pi^2)}{16\pi^2F^2} + \frac{A_0(m_K^2)}{32\pi^2F^2} \bigg] \,.
\end{eqnarray}
The one-point loop function $A_0(m^2)$ is calculated in dimensional regularization within the $\overline{MS}-1$ scheme~\cite{gasser8485} and it reads
\begin{equation}
 A_0(m^2)= -m^2 \ln{\dfrac{m^2}{\mu^2}}\,,
\end{equation}
with the renormalization scale $\mu$ fixed at $770$~MeV throughout.
 Using Eq.~\eqref{fpinlof0} to replace $F$ by $F_\pi$ in the NLO and NNLO
corrections, the resulting form is
\begin{eqnarray}\label{fpinnlofpi}
 F_\pi =&& F \bigg[ 1 + 4  L_5\frac{ m_\pi^2}{ F_\pi^2}
 + 4  L_4 \frac{ m_\pi^2 + 2m_K^2 }{ F_\pi^2}  +
 (56 L_5^2 - 64  L_5 L_8)\frac{m_\pi^4}{F_\pi^4}
 +  (8 C_{14} + 8 C_{17} )\frac{m_\pi^4}{F_\pi^2}
\nn \\ && \qquad
 + \frac{A_0(m_\pi^2)}{16\pi^2F_\pi^2} + \frac{A_0(m_K^2)}{32\pi^2F_\pi^2} \bigg]\,.
\end{eqnarray}
In the $\delta$ expansion, the expressions for a physical quantity with $F$ or $F_\pi$ in the higher order chiral corrections differ only for the $ L_5 L_{j=5,8}$ and
$ L_5\Lambda_{j=1,2}$ terms, since the differences by replacing $F$ by $F_\pi$ are originated from the NLO expressions of $F_\pi$ in Eq.~\eqref{fpinlofpi} and we only
retain terms up to NNLO in this work.
It is clear that the difference between Eqs.~\eqref{fpinnlof0} and \eqref{fpinnlofpi} is the $ L_5^2$ term.
Notice that in the $\delta$ expansion scheme, the terms like $ L_5 L_4$
are  N$^3$LO and will be dropped throughout the article.

The corresponding expression for the kaon decay constant
 when one uses $F$ to express the NLO and NNLO corrections reads
\begin{eqnarray}\label{fknnlof0}
 F_K =&& F \bigg[ 1 + 4  L_5\frac{ m_K^2}{ F^2}
 + 4  L_4 \frac{ m_\pi^2 + 2m_K^2 }{ F^2}  +
 (24 L_5^2 - 64  L_5 L_8)\frac{m_K^4}{F^4}
 +  8 C_{14} \frac{2m_K^4 - 2m_K^2m_\pi^2 +m_\pi^4}{F^2}
\nn \\ && \qquad
 +  8 C_{17} \frac{ m_\pi^2(2m_K^2 - m_\pi^2) }{F^2}
 + \frac{3A_0(m_\pi^2)}{128\pi^2F^2} + \frac{3A_0(m_K^2)}{64\pi^2F^2}
 + \frac{3c_\theta^2 A_0(m_\eta^2)}{128\pi^2F^2} + \frac{3s_\theta^2 A_0(m_{\eta'}^2)}{128\pi^2F^2}  \bigg]
 \,.
\end{eqnarray}
 On the other hand, expressing the NLO and NNLO contributions in terms of $F_\pi$ yields
\begin{eqnarray}\label{fknnlofpi}
 F_K =&& F \bigg[ 1 + 4  L_5\frac{ m_K^2}{ F_\pi^2}
 + 4  L_4 \frac{ m_\pi^2 + 2m_K^2 }{ F_\pi^2}
+ 8 L_5^2\frac{3m_K^4 + 4m_K^2m_\pi^2}{F_\pi^4}
 - 64  L_5 L_8\frac{m_K^4}{F_\pi^4}
\nn \\ && \qquad
 +  8 C_{14} \frac{2m_K^4 - 2m_K^2m_\pi^2 +m_\pi^4}{F_\pi^2}
 +  8 C_{17} \frac{ m_\pi^2(2m_K^2 - m_\pi^2) }{F_\pi^2}
 \nn \\ && \qquad
 + \frac{3A_0(m_\pi^2)}{128\pi^2F_\pi^2} + \frac{3A_0(m_K^2)}{64\pi^2F_\pi^2}
 + \frac{3c_\theta^2 A_0(m_\eta^2)}{128\pi^2F_\pi^2} + \frac{3s_\theta^2 A_0(m_{\eta'}^2)}{128\pi^2F_\pi^2}
 \bigg] \,.
\end{eqnarray}

The expanded expression for the ratio of $F_K/F_\pi$ in terms of $F$ up to NNLO in $\delta$ expansion, takes the form
\begin{eqnarray}\label{frinnlof0}
 \dfrac{F_K}{F_\pi} =&& 1 + 4  L_5\frac{ m_K^2-m_\pi^2}{ F^2}
 + 8  L_5^2\frac{3m_K^4-2m_K^2m_\pi^2-m_\pi^4}{F^4}
 +64 L_5L_8\dfrac{m_\pi^4-m_K^4}{F^4}
 \nn \\ &&
 +16 C_{14}\dfrac{m_K^4-m_K^2m_\pi^2}{F^2}
 +16 C_{17}\dfrac{m_K^2m_\pi^2-m_\pi^4}{F^2}
 \nn \\ &&
 -\dfrac{5A_0(m_\pi^2)}{128\pi^2F^2}+\dfrac{A_0(m_K^2)}{64\pi^2F^2}
 +\dfrac{3c_\theta^2 A_0(m_\eta^2)}{128\pi^2F^2}+\dfrac{3s_\theta^2 A_0(m_{\eta'}^2)}{128\pi^2F^2}
 \,.
\end{eqnarray}
When expressing the previous result in terms of $F_\pi$, it reads
\begin{eqnarray}\label{frinnlofpi}
 \dfrac{F_K}{F_\pi} =&& 1 + 4  L_5\frac{ m_K^2-m_\pi^2}{ F_\pi^2}
 + 8  L_5^2\frac{3m_K^4+2m_K^2m_\pi^2-5 m_\pi^4}{F_\pi^4}
 +64 L_5L_8\dfrac{m_\pi^4-m_K^4}{F_\pi^4}
 \nn \\ &&
 +16 C_{14}\dfrac{m_K^4-m_K^2m_\pi^2}{F_\pi^2}
 +16 C_{17}\dfrac{m_K^2m_\pi^2-m_\pi^4}{F_\pi^2}
 \nn \\ &&
 -\dfrac{5A_0(m_\pi^2)}{128\pi^2F_\pi^2}+\dfrac{A_0(m_K^2)}{64\pi^2F_\pi^2}
 +\dfrac{3c_\theta^2 A_0(m_\eta^2)}{128\pi^2F_\pi^2}+\dfrac{3s_\theta^2 A_0(m_{\eta'}^2)}{128\pi^2F_\pi^2}
 \,,
\end{eqnarray}
which differs from Eq.~\eqref{frinnlof0} in the $L_5^2$ term.

The pion squared mass up to NNLO is given by
\begin{eqnarray}\label{mpi2mpi02}
m^{2}_{\pi} =&& \overline{m}_\pi^2 + m_\pi^{2,{\rm NLO}} + m_\pi^{2,{\rm NNLO}}\,,
\end{eqnarray}
with
\begin{eqnarray}
\overline{m}_\pi^2  =&& 2 B \widehat{m} \, ,
\\
\label{mpi2nlof0}
m_\pi^{2,{\rm NLO}} =&& \dfrac{8(2 L_8- L_{5})m_{\pi}^{4}}{F^{2}}\,, \\
 m_\pi^{2,{\rm NNLO}}= && \dfrac{8(2 L_6- L_{4})m_{\pi}^{2}(2m_K^2+m_\pi^2)}{F^{2}}
-\frac{ 64( L_5^2- 6 L_5 L_8+ 8 L_8^2)m_\pi^6}{F^4}
\nn \\ &&
-\frac{16(2C_{12}+C_{14}+C_{17}-3C_{19}-2C_{31})m_\pi^6}{F^2}
+ \dfrac{m_{\pi}^{2}(c_\theta^{2}-2\sqrt{2}c_\theta s_\theta+2s_\theta^{2})A_{0}(m_{\eta}^{2})}{96\pi^2F^{2}}
\nn \\ &&
+\dfrac{m_{\pi}^{2}(2c_\theta^{2}+2\sqrt{2}c_\theta s_\theta+s_\theta^{2})A_{0}(m_{\eta^{'}}^{2})}{96\pi^2F^{2}}
-\dfrac{m_{\pi}^{2}A_{0}(m_{\pi}^{2})}{32\pi^2F^{2}}\,. \label{mpi2nnlof0}
\end{eqnarray}
When expressing the renormalized $m_\pi$ in terms of $F_\pi$, the only differences are the $L_5L_8$ and $L_5^2$ terms in Eq.~\eqref{mpi2nnlof0}
and the other parts are the same as in Eq.~\eqref{mpi2mpi02} with the explicit replacement of $F$ by $F_\pi$ in Eq.~\eqref{fpinlof0}.
Therefore we only give the different parts for simplicity when expressing in terms of $F_\pi$ and they read
\begin{eqnarray}
 m_\pi^{2,{(F_\pi)},\,L_5L_8\,,L_5^2}=&&
\frac{ 128(4L_5 L_8- L_5^2 )m_\pi^6}{F_\pi^4}\,.
\end{eqnarray}

The mass squared for kaon up to NNLO is  provided by
\begin{eqnarray}\label{mk2mk02}
m^{2}_{K} =&& \overline{m}_K^2 + m_K^{2,{\rm NLO}} + m_K^{2,{\rm NNLO}}
\end{eqnarray}
with
\begin{eqnarray}
  \overline{m}_K^2    =&&   B(\widehat{m}+m_s)\, ,
\\
m_K^{2,{\rm NLO}}= && \dfrac{8(2 L_8- L_{5})m_{K}^{4}}{F^{2}}   \, ,
\\
m_K^{2,{\rm NLO}}=&& \dfrac{8(2 L_6- L_{4})m_{K}^{2}(2m_K^2+m_\pi^2)}{F^{2}}
-\frac{ 64( L_5^2- 6 L_5 L_8+ 8 L_8^2)m_K^6}{F^4}
\nn \\ &&
-\dfrac{32C_{12}m_{K}^{6}}{F^{4}}+\dfrac{32C_{31}m_{K}^{6}}{F^{2}}
+\dfrac{16C_{17}m_{K}^{2}m_{\pi}^{2}(-2m_{K}^{2}+m_{\pi}^{2})}{F^{2}}
-\dfrac{16C_{14}m_{K}^{2}(2m_{K}^{4}-2m_{K}^{2}m_{\pi}^{2}+m_{\pi}^{4})}{F^{2}} 
\nn \\ &&
+\dfrac{48C_{19}m_{K}^{2}(2m_{K}^{4}-2m_{K}^{2}m_{\pi}^{2}+m_{\pi}^{4})}{F^{2}}
-\dfrac{[c_\theta^{2}(3m_{\eta}^{2}+m_{\pi}^{2})+2\sqrt{2}c_\theta s_\theta(-2m_{K}^{2}+m_{\pi}^{2})
-4m_{K}^{2}s_\theta^{2}]A_{0}(m_{\eta}^{2})}{192\pi^2F^{2}}
\nn \\ &&
-\dfrac{[-4c_\theta^{2}m_{K}^{2}+2\sqrt{2}c_\theta s_\theta(2m_{K}^{2}-m_{\pi}^{2})
+(3m_{\eta^{'}}^{2}+m_{\pi}^{2})s_\theta^{2}]A_{0}(m_{\eta^{'}}^{2})}{192\pi^2F^{2}}\,.  \label{mk2nnlof0}
\end{eqnarray}
When expressing Eq.~\eqref{mk2mk02} in terms of $F_\pi$, the differences are the $L_5L_8$ and $L_5^2$ terms
in Eq.~\eqref{mk2nnlof0} and the new expressions are
\begin{eqnarray}
 m_K^{2,{(F_\pi)},\,L_5L_8\,,L_5^2}=&&
-\frac{ 64L_5^2 m_K^4(m_K^2+m_\pi^2)}{F_\pi^4}+\frac{ 128L_5 L_8 m_K^4(3m_K^2+m_\pi^2)}{F_\pi^4}\,.
\end{eqnarray}
Notice that the masses of pion and kaon appearing in NLO and NNLO parts in
the above equations correspond to the renormalized quantities, instead of their LO expressions.
In addition, this gives the quark mass ratio relation  $m_s/\hat{m}= 2\overline{m}_K^2/\overline{m}_\pi^2 \,-\, 1$.

When performing the chiral extrapolation of the lattice data, instead of the renormalized $m_K^2$ as in the previous equations,
it is convenient to use the LO kaon mass squared in the higher order corrections. In this way, we do not need to
iteratively solve Eq.~\eqref{mk2mk02} in order to give the value of $m_K$ for a given $m_\pi$. The result in terms
of $\overline{m}_K$ in the NLO and NNLO expressions becomes
\begin{eqnarray}\label{mk2mk02bar}
m^{2,{\rm Lat}}_{K} =&& \overline{m}_K^2  + m_K^{2,{\rm Lat-NLO}}  + m_K^{2,{\rm Lat-NNLO}}\,,
\end{eqnarray}
with
\begin{eqnarray}
 m_K^{2,{\rm Lat-NLO}} =&& \dfrac{8(2 L_8- L_{5})\overline{m}_{K}^{4}}{F^{2}}\,, \\
 m_K^{2,{\rm Lat-NNLO}} =&&
  \dfrac{8(2 L_6- L_{4})\overline{m}_{K}^{2}(2\overline{m}_K^2+m_\pi^2)}{F^{2}}
+\frac{ 64( L_5^2- 2 L_5 L_8)\overline{m}_K^6}{F^4}
-\dfrac{32C_{12}\overline{m}_{K}^{6}}{F^{4}}+\dfrac{32C_{31}\overline{m}_{K}^{6}}{F^{2}}
\nn \\ &&
+\dfrac{16C_{17}\overline{m}_{K}^{2}m_{\pi}^{2}(-2\overline{m}_{K}^{2}+m_{\pi}^{2})}{F^{2}}
-\dfrac{16C_{14}\overline{m}_{K}^{2}(2\overline{m}_{K}^{4}-2\overline{m}_{K}^{2}m_{\pi}^{2}+m_{\pi}^{4})}{F^{2}}
\nn \\ &&
+\dfrac{48C_{19}\overline{m}_{K}^{2}(2\overline{m}_{K}^{4}-2\overline{m}_{K}^{2}m_{\pi}^{2}
+m_{\pi}^{4})}{F^{2}}
\nn \\ &&
-\dfrac{[c_\theta^{2}(3m_{\eta}^{2}+m_{\pi}^{2})+2\sqrt{2}c_\theta s_\theta(-2\overline{m}_{K}^{2}+m_{\pi}^{2})
-4\overline{m}_{K}^{2}s_\theta^{2}]A_{0}(m_{\eta}^{2})}{192\pi^2F^{2}}
\nn \\ &&
-\dfrac{[-4c_\theta^{2}\overline{m}_{K}^{2}+2\sqrt{2}c_\theta s_\theta(2\overline{m}_{K}^{2}-m_{\pi}^{2})
+(3m_{\eta^{'}}^{2}+m_{\pi}^{2})s_\theta^{2}]A_{0}(m_{\eta^{'}}^{2})}{192\pi^2F^{2}}\,. \label{mk2nnlomkbarf0}
\end{eqnarray}
When expressing Eq.~\eqref{mk2mk02bar} in terms of $F_\pi$, the differences are the $L_5L_8$ and $L_5^2$ terms
in Eq.~\eqref{mk2nnlomkbarf0} and the new expressions are
\begin{eqnarray}
 m_K^{2,{\rm Lat},{(F_\pi)},\,L_5L_8\,,L_5^2}=&&
\frac{ 64L_5 (L_5-2L_8) \overline{m}_K^4(\overline{m}_K^2-m_\pi^2)}{F_\pi^4}\,.
\end{eqnarray}

When confronting with the lattice data,  we only consider the simulated points with the physical
strange-quark mass, i.e. the lattice ensembles that when extrapolating to the
physical pion masses lead simultaneously to physical kaon masses.
In this case, we can express the LO kaon mass squared as
\begin{eqnarray}
 \overline{m}_K^2= B(m_s^{\rm Phy} + \widehat{m}   )
 =  B(m_s^{\rm Phy}+ \widehat{m}^{\rm Phy}) - B\widehat{m}^{\rm Phy}  + B\widehat{m}
 = \overline{m}_K^{2,{\rm Phy}}- \frac{\overline{m}_\pi^{2,{\rm Phy}}}{2} + \frac{\overline{m}_\pi^{2}}{2}  \,,
\end{eqnarray}
where $\overline{m}_K^{2,{\rm Phy}}$ and $\overline{m}_\pi^{2,{\rm Phy}}$ can be obtained
through Eqs.~\eqref{mpi2mpi02} and \eqref{mk2mk02} by substituting the physical
masses of $\pi,K,\eta,\eta'$ in the NLO and NNLO expressions.
For $\overline{m}_\pi^{2}$, which varies in the lattice
simulation, we can extract its value by using Eq.~\eqref{mpi2mpi02}. In this case,
$m_\pi$ in Eq.~\eqref{mpi2mpi02} takes the value from lattice simulation,
and $m_K,m_\eta,m_{\eta'}$, which only appear in the NNLO part, can be approximated
by their LO expressions.

In the above discussions, we have distinguished the situations of using $1/F^2$ and $1/F_\pi^2$ in the higher order corrections for 
various observables. Similarly, we can also generalize the discussions by replacing the renormalized masses ($m_\pi$ and $m_K$)  with  
the LO ones ($\overline{m}_\pi$ and $\overline{m}_K$) in the higher order corrections. 
We take the observables $F_\pi$  and $F_K$ as examples to illustrate the differences. 
The renormalized $m_\pi$ and $m_K$ have been used in Eqs.~\eqref{fpinnlof0} and \eqref{fknnlof0} for $F_\pi$ and $F_K$ with $1/F^2$ in the 
higher order terms, respectively.  After replacing $m_\pi$ and $m_K$ in Eqs.~\eqref{fpinnlof0} and \eqref{fknnlof0} 
with their expressions in terms of the LO masses $\overline{m}_\pi$ and $\overline{m}_K$ 
through Eqs.~\eqref{mpi2mpi02} and \eqref{mk2mk02} respectively, the corresponding expressions 
are found to be
\begin{eqnarray}\label{fpinnlof0mass0}
 F_\pi =&& F \bigg[ 1 + 4  L_5\frac{ \overline{m}_\pi^2}{ F^2}
 + 4  L_4 \frac{ \overline{m}_\pi^2 + 2\overline{m}_K^2 }{ F^2}  
 -8L_5^2\frac{\overline{m}_\pi^4}{F^4}
 +  (8 C_{14} + 8 C_{17} )\frac{\overline{m}_\pi^4}{F^2}
 + \frac{A_0(\overline{m}_\pi^2)}{16\pi^2F^2} + \frac{A_0(\overline{m}_K^2)}{32\pi^2F^2} \bigg] \,,
\end{eqnarray}
\begin{eqnarray}\label{fknnlof0mass0}
 F_K =&& F \bigg[ 1 + 4  L_5\frac{ \overline{m}_K^2}{ F^2}
 + 4  L_4 \frac{ \overline{m}_\pi^2 + 2\overline{m}_K^2 }{ F^2}  
- 8L_5^2 \frac{\overline{m}_K^4}{F^4}
 +  8 C_{14} \frac{2\overline{m}_K^4 - 2\overline{m}_K^2\overline{m}_\pi^2 +\overline{m}_\pi^4}{F^2}
\nn \\ && \qquad
 +  8 C_{17} \frac{ \overline{m}_\pi^2(2\overline{m}_K^2 - \overline{m}_\pi^2) }{F^2}
 + \frac{3A_0(\overline{m}_\pi^2)}{128\pi^2F^2} + \frac{3A_0(\overline{m}_K^2)}{64\pi^2F^2}
 + \frac{3c_\theta^2 A_0(\overline{m}_\eta^2)}{128\pi^2F^2} + \frac{3s_\theta^2 A_0(\overline{m}_{\eta'}^2)}{128\pi^2F^2}  \bigg]
 \,.
\end{eqnarray}
As in the discussion of $1/F^2$ versus $1/F_\pi^2$ up to the NNLO precision, the expressions for a specific observable by using the 
renormalized masses $m_\pi, m_K$ and the LO $\overline{m}_\pi, \overline{m}_K$ only differ in 
the terms like $L_i L_j$, being $L_i$ and $L_j$ the NLO LECs in Eq.~\eqref{lagnlo}. This can be clearly seen when 
comparing Eqs.~\eqref{fpinnlof0} and \eqref{fpinnlof0mass0}. E.g. the differences caused by using the renormalized 
masses and the LO ones are the $L_5^2$ and $L_5L_8$ terms, apart from the explicit replacement of $m_\pi$ and $m_K$ 
by $\overline{m}_\pi$ and $\overline{m}_K$ respectively. Similar rules are also applied to Eqs.~\eqref{fknnlof0} and \eqref{fknnlof0mass0}. 

To replace $m_\pi, m_K$ by $\overline{m}_\pi$ and $\overline{m}_K$ in the NLO and NNLO corrections in Eq.~\eqref{mpi2mpi02}, the only changes happen for 
the $L_i L_j$ terms and the corresponding new expressions read 
\begin{eqnarray}
 m_\pi^{2,{(\overline{m}_\pi,\overline{m}_K, F)},\,L_5L_8\,,L_5^2\,,L_8^2}=&&
\frac{ 64L_5(L_5- 2L_8)\overline{m}_\pi^6}{F^4}\,.
\end{eqnarray}

In principle, we should also present the results expressed with the LO masses $\overline{m}_\pi, \overline{m}_K$ and the renormalized decay constants 
$F_\pi, F_K$, which can be straightforwardly obtained by substituting the relations in Eqs.~\eqref{mpi2mpi02} and \eqref{mk2mk02} into the corresponding 
observables. We consider the expressions given in terms of the renormalized masses and $1/F^2$ as our preferred ones in this work. 
The reason to choose the renormalized masses is for practical purpose, 
since in lattice simulations the different observables are typically given as functions of the renormalized $m_\pi^2$. 
Also most of the chiral studies choose to express the quantities with the renormalized masses in the higher order corrections, such as in
Refs.~\cite{Amoros:1999dp,Bernard:2009ds}. Following this rule we consider the results with the renormalized masses and $1/F_\pi^2$ as an estimate 
of systematic errors due to the truncation of the $\delta$ expansion when one works at a given order in perturbation theory.
While for the case with the LO masses, we shall also comment the results in the following numerical discussions.

\section{Phenomenological discussions}\label{sec.pheno}

The big challenge in the present general discussions on the $\eta$-$\eta'$ mixing is the determination of the unknown
LECs in Eqs.~\eqref{laglo},~\eqref{lagnlo} and \eqref{lagnnlo}.  The recent lattice simulations on the light
pseudoscalar mesons provide us valuable sources to constrain these free parameters.
The considered lattice simulations include the $m_\pi$ dependences of the masses of $\eta, \eta'$~\cite{etm13prl,etm13pos,ukqcd12prd,rbcukqcd10prl,hsc11prd}
and kaon~\cite{rbcukqcd11prd,rbcukqcd13prd}, and the $\pi, K$ decay constants~\cite{rbcukqcd11prd,rbcukqcd13prd} and their ratios~\cite{Durr:2010hr}.
Moreover, relevant phenomenological results and experimental data
will be also included to constrain the LECs.

Since we do not consider the isospin violating effects, we will take the values for the physical pion and kaon masses
in the isospin limit from Ref.~\cite{Aoki:2013ldr}, where the corrections from the electromagnetic contributions are removed,
\begin{equation}
 m_\pi= 135.0~{\rm MeV}\,, \qquad  m_K= 494.2~{\rm MeV}\, .
\end{equation}
These values will be used in later chiral extrapolations, while  for the physical masses of $\eta$ and $\eta'$ and
the decay constants of pion and kaon, we will take their world-average values from Ref.~\cite{pdg14}.

In order to show the results step by step, we present the discussions
in the following sections split in three parts: we consider fits performed
at leading order, next-to-leading order and next-to-next-to-leading order.

\subsection{Leading-order analyses}\label{sectlo}

At leading order, the $\eta$-$\eta'$ mixing is described by one free parameter, namely the singlet $\eta_0$ mass $M_0$ in Eq.~\eqref{laglo}
and the explicit expressions for the masses and mixing angle are given in Eqs.~\eqref{defmetab2},~\eqref{defmetaPb2} and \eqref{deftheta0}.
At this order, the $\pi, K$ decay constants are degenerate and
given by their chiral and large $N_C$ limits, i.e. $F_\pi=F_K=F$.
Therefore we shall not take the lattice simulations of the decay constants into account for the LO discussion
as they clearly show the need of higher order corrections for a suitable description.
Also at leading order, $F$ will not enter the masses and mixing angle, as shown in Eqs.~\eqref{defmetab2},~\eqref{defmetaPb2} and \eqref{deftheta0}.
As a result of this,
we do not need to distinguish the two situations with $F$ or $F_\pi$
discussed previously  in the expressions of different observables.
Apart from the lattice simulation data, we also fit the physical values of the $\eta$ and $\eta'$ masses.
 Nonetheless, fitting   the physical masses with the experimental
 precision at the level of several hundred-thousandth is too ambitious.
Since the ultimate goal of the present work is the NNLO study, the ballpark estimate of our theoretical uncertainty, starting from the N$^3$LO part,
should be around 3\%. This value is obtained from the general rule that each higher order correction in $\delta$ expansion,
either the $SU(3)$-flavor breaking or the $1/N_C$ effect, is around $30\%$. In fact, the estimated three-percent uncertainty is also
similar to the typical error bars reported in many lattice simulations,
 in the range from $3-10\%$~\cite{etm13prl,etm13pos,ukqcd12prd,rbcukqcd10prl,hsc11prd}.
Consistently, we assign a 1\%~uncertainty to the physical values of $m_\eta$ and $m_{\eta'}$ in the fits.

\begin{figure}[ht]
\begin{center}
 \includegraphics[angle=0.0,width=1.0\textwidth]{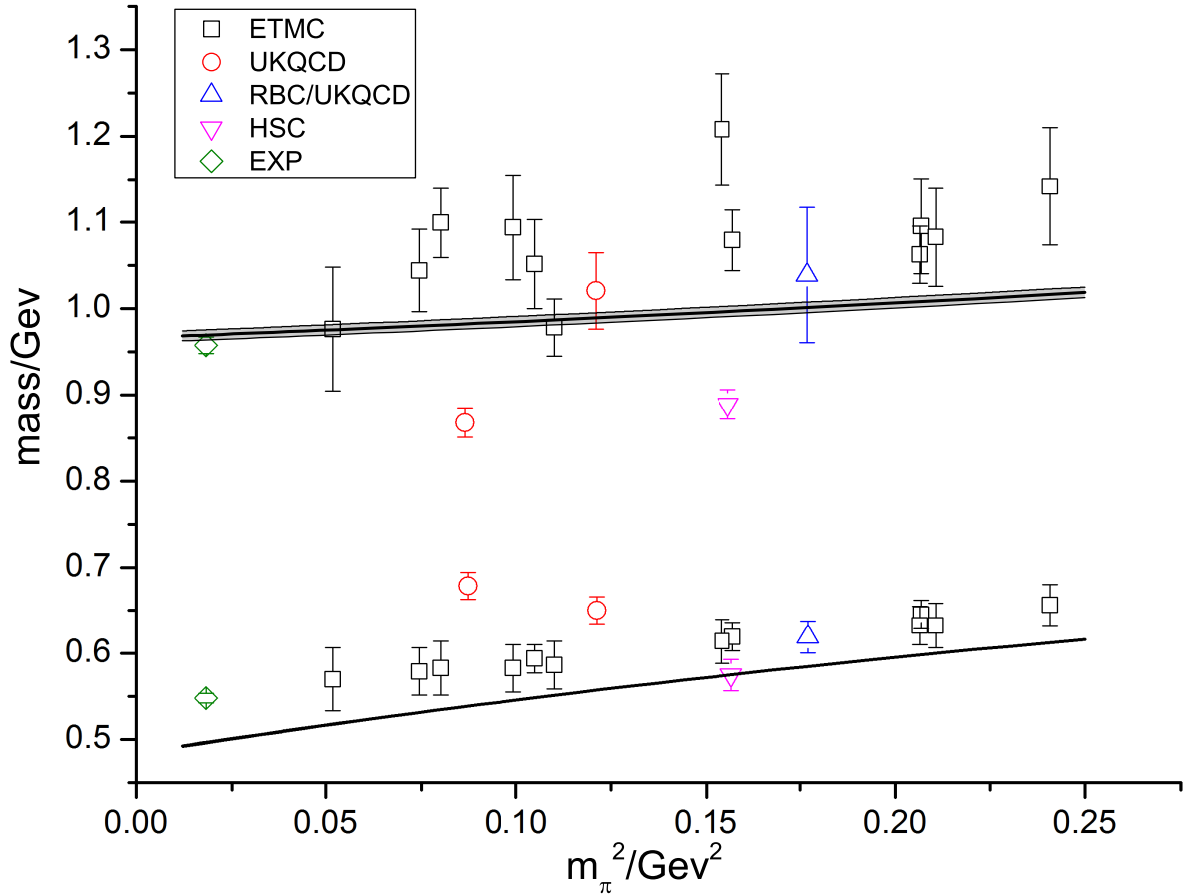}
\caption{The masses of $\eta$ and $\eta'$ from the LO fit. The left most two points correspond to the physical masses.
The remaining lattice simulation data are taken from Refs.~\cite{etm13pos,etm13prl}~(ETMC), \cite{ukqcd12prd}~(UKQCD),
\cite{rbcukqcd10prl}~(RBC/UKQCD), \cite{hsc11prd}~(HSC), where we only take into account the simulation points with $m_\pi<500$~MeV.
The shade area surrounding each curve stands for the statistical uncertainty from the fit. }\label{figlometa}
 \end{center}
\end{figure}

The value of the singlet mass $M_0$ from the LO fit is
\begin{equation}\label{lofit}
 M_0= ( 835.7 \pm 7.5 )\,{\rm MeV} \,.
\end{equation}
The physical masses for the $\eta, \eta'$ and their LO mixing angle $\theta$ from the fit are found to be
\begin{equation}
 m_\eta= (496.4 \pm 1.3) \, {\rm MeV}\,, \quad  m_{\eta'}= (969.8 \pm 5.8 ) \, {\rm MeV}\,, \theta= -18.9^{\circ} \pm 0.3^\circ\,.
\end{equation}
The resulting plots can be seen in Fig.~\ref{figlometa}.  We verify that if the physical masses are excluded in the fit, $M_0=813 \pm 11 $~MeV results.
If we only include the physical masses and exclude the lattice simulation data in the fit, $M_0=859 \pm 11$~MeV is obtained.
These determinations of $M_0$ lie within the broad range summarized in Ref.~\cite{Feldmann:1999uf} and are quite close with the commonly used
values of $M_0=850$~MeV~\cite{Feldmann:1999uf}. Taking into account the large uncertainties of the lattice simulation data, specially
for $m_{\eta'}$, and the concise formalism of the LO mixing, it is impressive that the lattice simulation data
 can already be qualitatively described with the LO analysis, as shown in Fig.~\ref{figlometa}.
This also indicates that the higher order mixing effects can only give moderate corrections to the masses of $\eta$ and $\eta'$.

Nevertheless, in order to describe the lattice data more accurately, specially the $\eta$ masses,
the chiral corrections beyond the leading order are needed.
For the physical masses, it has also been shown that the LO description fails to explain the mass ratio
of $\eta$ and $\eta'$  accurately enough~\cite{Georgi:1993jn}.
Therefore it is essential to generalize the discussions to NLO and NNLO
in order to achieve a precise description both for lattice simulations and physical data.

\subsection{ Next-to-leading order analyses}\label{sectnlo}

At next-to-leading order,
in addition to  the parameter $M_0$ at leading order, there are five additional free parameters: the decay constant $F$
at chiral and large $N_C$ limits, and the four NLO LECs $ L_5$, $ L_8$, $\Lambda_1$ and $\Lambda_2$ in Eq.~\eqref{lagnlo}.
At this order, as well as at next-to-next-to-leading order,
 one can rewrite the chiral expansion of the observables in various equivalent ways up to the perturbative
order in $\delta$ under consideration.
In the following discussion, we will perform
two types of fits: one using $F$
in the theoretical NLO and NNLO expressions and the other employing $F_\pi$,
as discussed in Sect.~\ref{sect.nnloformula}.
Since the differences of the theoretical expressions used in the two types of fits are beyond the considered precision,
the variances of the outputs from the two fits can be considered as systematic errors from the theoretical models by neglecting higher order contributions.
In the following, we will explicitly present the fit results by using $F$ in the theoretical expressions, which is the most straightforward option, as discussed in
Sect.~\ref{sect.nnloformula}. The outputs of the fits  with the theoretical formulas expressed in terms of $F_\pi$  will be used to estimate the systematic errors:
the difference between the central values of the two types of fits will be used to estimate the truncation uncertainty due to
working just up to a given order in the $\delta$--expansion, providing the second error for each quantity in the following tables.

In Refs.~\cite{Gerard:2004gx,Gerard:2009ps,Mathieu:2010ss}, it is argued that at each chiral order, the leading $N_C$ effects
are dominant, or in other words that the $\Lambda_1$ and $\Lambda_2$ terms are assumed to be much less irrelevant than the $ L_5$
and $ L_8$ terms in the NLO $\delta$ expansion. This assumption has been more or less confirmed when focusing on the masses of $\eta, \eta'$ and
the LO mixing angle at the physical points~\cite{Gerard:2004gx,Gerard:2009ps,Mathieu:2010ss}.
In Ref.~\cite{Guo:2011pa}, the local higher order LECs were estimated by the tree-level resonance exchanges
and it was found that with those LECs $\Lambda_2$ seems to be more important than $\Lambda_1$ when focusing on
the physical masses for $\eta$ and $\eta'$.
It is interesting to check how these assumptions work when including the lattice simulations and the phenomenological
results of the two-mixing-angle parameters, which are not considered in Refs.~\cite{Gerard:2004gx,Gerard:2009ps,Mathieu:2010ss,Guo:2011pa}.
Different sets of fits to the lattice data and phenomenological inputs from the two-mixing-angle scheme are performed either
by fixing $\Lambda_{i=1,2}$ to zero or releasing their values, in order to reexamine the assumptions.
Interestingly we do not find qualitative changes between the fits
with fixed $\Lambda_{i=1,2}=0$ and the ones with free values for these parameters.
This tells us that indeed the $\Lambda_1$ and $\Lambda_2$ terms
do not significantly improve the fit results, even after taking into account the lattice simulations.
Nevertheless, we find that these two terms are quite important to reproduce the phenomenological mixing
angles $\theta_0$ and $\theta_8$ in the fits where $M_0$ is fixed
at its LO value. If $M_0$ is released in the fits we find that including
$\Lambda_1$ and $\Lambda_2$  improves the descriptions of $m_{\eta'}$ from lattice simulations.
Therefore, we will not further discuss fits with $\Lambda_1$ and $\Lambda_2$ set to
zero in the following. Instead, we focus on the results given in Table~\ref{tabnlofit}
 with all   the four NLO LECs in the fits, namely $ L_5$, $ L_8$, $\Lambda_1$ and $\Lambda_2$ in Eq.~\eqref{lagnlo}.

For the parameter $M_0$, we take two strategies to estimate its value in NLO analysis.
In one of them we fix $M_0=835.7$~MeV from its LO determination (NLOFit-A)
and in the other case we free its value for the NLO fit (NLOFit-B).
These two NLO fits are given in Table~\ref{tabnlofit}.
The first error bar for each fitted parameter corresponds to the statistical one from the fits and the second error bar is estimated
from the variation of the fits between those using $F$ and $F_\pi$ in the
 NLO (and later also NNLO)
theoretical expressions.
From the two fits shown in Table~\ref{tabnlofit}, one can see that releasing $M_0$ in the fits barely changes
the fit quality with respect to
the cases when its value is fixed,
although there are slight variations in the determinations of $M_0$ and $\Lambda_2$.

Concerning the results of the LECs in Table~\ref{tabnlofit}, the resulting values for $F$ from the two fits are quite
compatible and close to the physical pion decay constant.
For $\Lambda_1$ and $\Lambda_2$, their values are poorly known in literature and it is helpful to
compare our values with the following estimate for their ranges:
we take the LO determination $M_0=835.7$~MeV, and we
then separately include the $\Lambda_1$ and $\Lambda_2$ terms
in the $\eta$-$\eta'$ mixing and vary their values to obtain new results for $m_\eta$ and $m_{\eta'}$ with the physical $m_\pi$. Since the $\Lambda_1$ and
$\Lambda_2$ terms are NLO $1/N_C$ effects, it is reasonable to assume that their corrections to $m_\eta^2$ or $m_{\eta'}^2$ should be at most
around $30\%$ of the LO results. In this way we can set up conservative and rough estimates for the ranges of $\Lambda_1$ and $\Lambda_2$, which
are found to be
\begin{eqnarray}\label{lam12range}
 |\Lambda_1|  < 0.4\,, \qquad  |\Lambda_2|< 0.7 \,\,\,.
\end{eqnarray}

The resulting magnitudes of $\Lambda_1$ in our fits are tiny and consistent with zero,
 as shown in Table~\ref{tabnlofit}.
For the parameter $\Lambda_2$, our determinations lie within the ranges estimated in Eq.~\eqref{lam12range}.
Its value, specially the one from NLOFit-A, is close to the one used in Ref.~\cite{Escribano:2010wt},
where the mixing was discussed at next-to-leading order. 
However the determinations for $\Lambda_2$ in Table~\ref{tabnlofit} become much more precise than those 
given in Refs.~\cite{Guo:2011pa,Guo:2012yt}, where the lattice simulations for $m_\eta$ and $m_{\eta'}$ are not included, indicating 
the usefulness of incorporating the lattice data in the $U(3)$ $\chi$PT study.   
Our determinations of $ L_5$ and $ L_8$ are in good agreement with the
leading  $N_C$ predictions from resonance chiral theory~\cite{Ecker89}, the $SU(3)$ one-loop results in Ref.~\cite{gasser8485}
 and the one-loop resonance chiral theory determination for $L_8$~\cite{L8-rcht}.
But the values here are clearly larger than those from the recent two-loop determinations~\cite{Bijnens:2011tb,Bijnens:2014lea},
the results from $K\pi$ scattering in the scalar channels~\cite{Jamin:2000wn},
 and the one-loop resonance chiral theory estimates for $L_5$~\cite{Guo:2014yva}.
The discrepancies of $ L_5$ and $ L_8$, comparing with the recent two-loop determinations~\cite{Bijnens:2011tb,Bijnens:2014lea},
can be eliminated once the $\cO(p^6)$ LECs are taken into account, as we will show in the NNLO discussion.

 The values of the parameters in the two-mixing-angle scheme and the mass ratio of strange and up/down quarks resulting from the fits are
given in Table~\ref{tabnloresult}. Similarly, the first error bar for each quantity is the statistical error and the second one corresponds
to the systematic error, which is obtained in the same way as the one in Table~\ref{tabnlofit}.
Notice that these inputs have already been satisfactorily reproduced in NLO analyses.

\begin{figure}[ht]
\begin{center}
 \includegraphics[angle=0.0,width=1.0\textwidth]{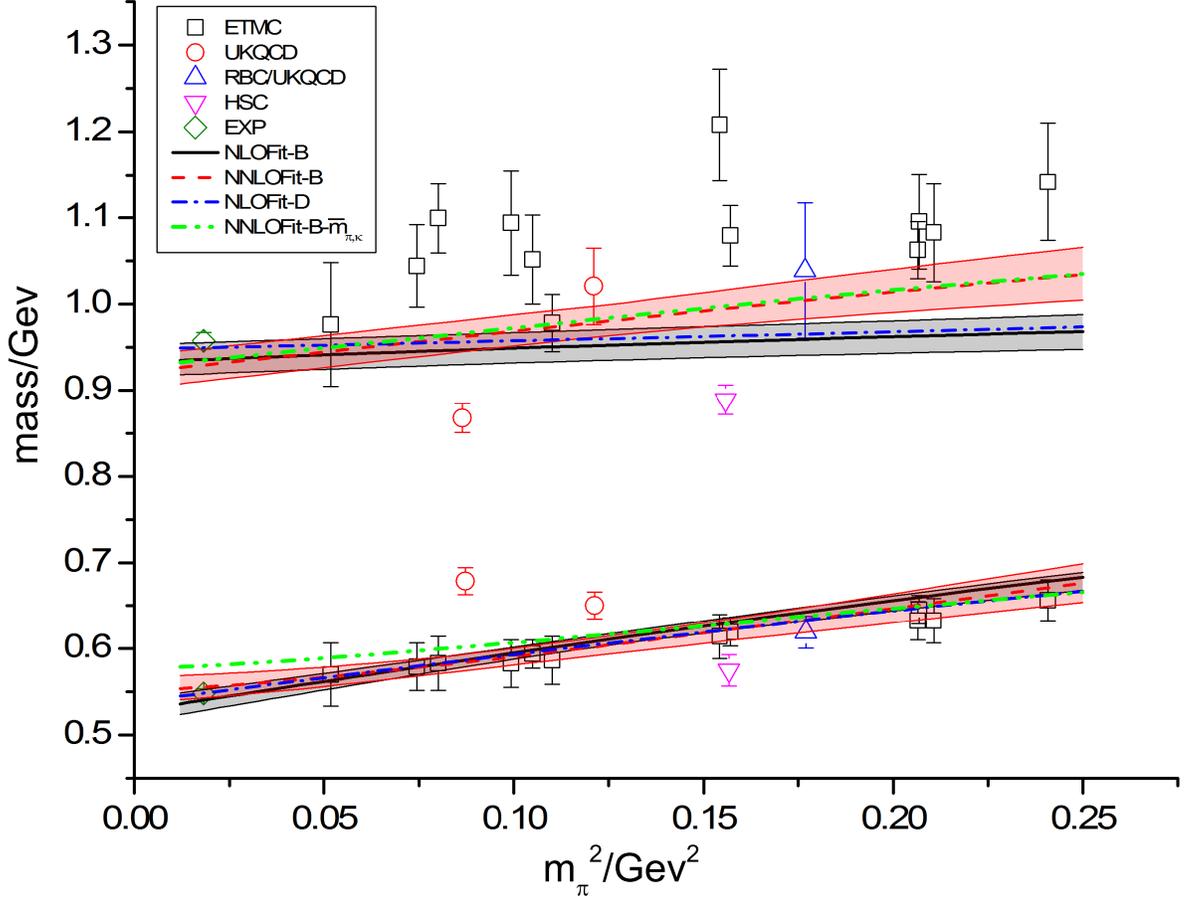}
\caption{The masses of $\eta$ and $\eta'$ from the NLO and NNLO fits. The left most two points correspond to the physical masses.
The remaining lattice simulation data are taken from Refs.~\cite{etm13pos,etm13prl}~(ETMC), \cite{ukqcd12prd}~(UKQCD),
\cite{rbcukqcd10prl}~(RBC-UKQCD), \cite{hsc11prd}~(HSC), where we only take into account the points with $m_\pi<500$~MeV.
The shaded areas around the black solid and red dashed lines stand for the statistical error bands from the NLOFit-B and NNLOFit-B fits, 
respectively. The meaning of notations for different lines are explained in detail in the text. 
}\label{figmeta}
 \end{center}
\end{figure}

\begin{figure}[ht]
\begin{center}
 \includegraphics[angle=0.0,width=1.0\textwidth]{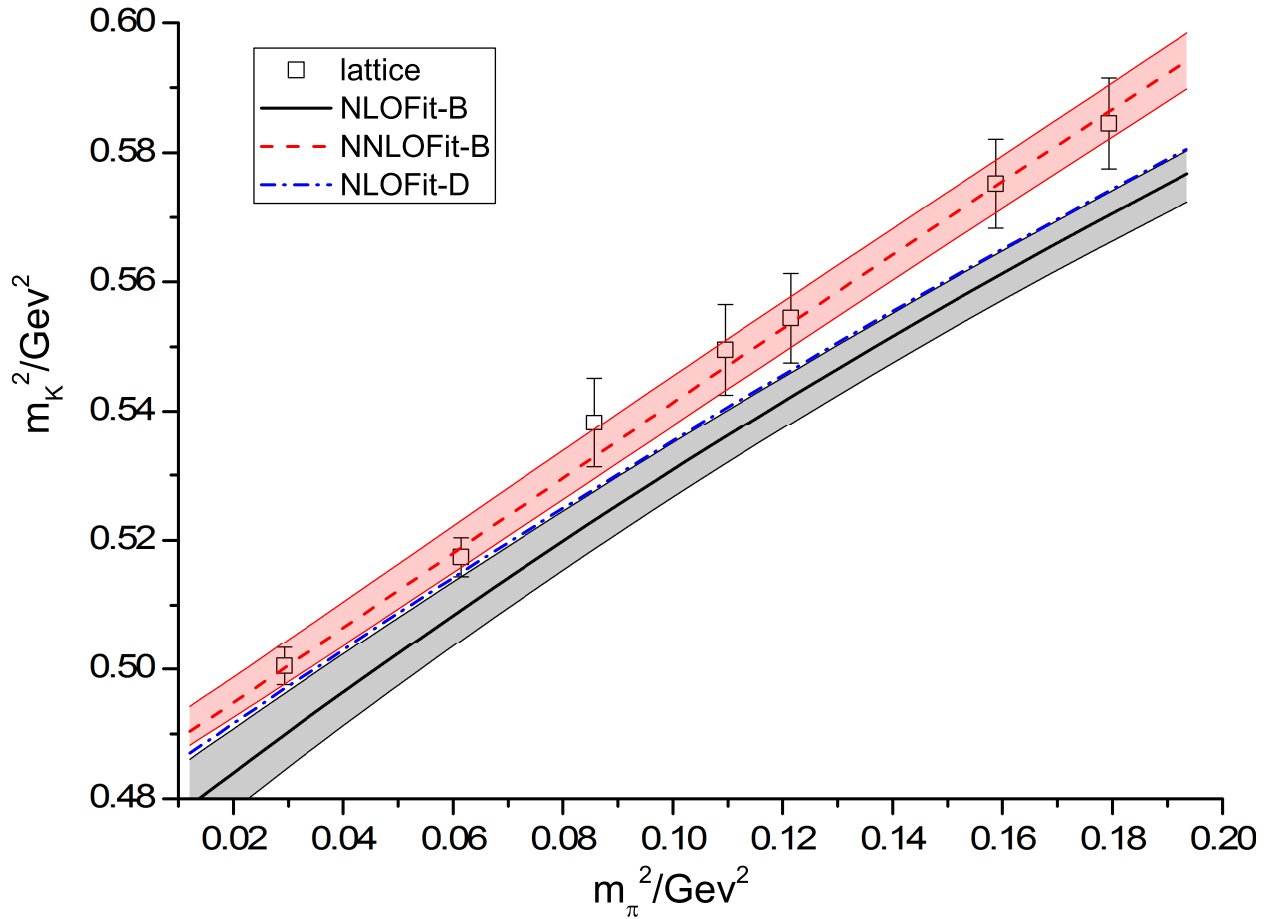}
\caption{  Kaon mass   from the NLO and NNLO fits. The lattice simulation data are taken from RBC and UKQCD~\cite{rbcukqcd11prd,rbcukqcd13prd}.
Only the unitary points simulated with the physical strange quark mass are included.
The shaded areas around the black solid and red dashed lines stand for the statistical error bands from the NLOFit-B and NNLOFit-B fits, 
respectively. 
The meaning of notations for different lines are explained in detail in the text. 
}\label{figmk}
 \end{center}
\end{figure}

\begin{figure}[ht]
\begin{center}
 \includegraphics[angle=0.0,width=1.0\textwidth]{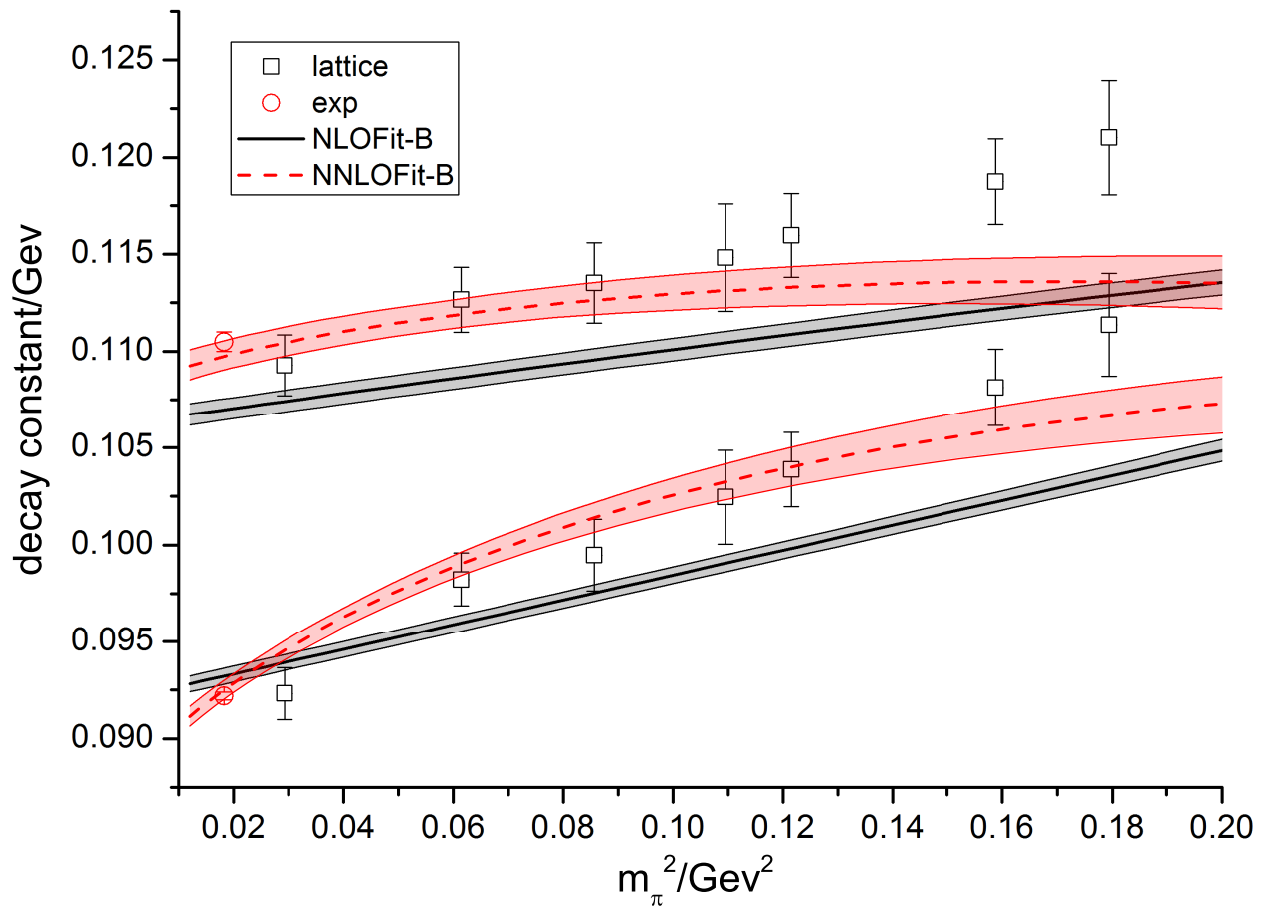}
\caption{ Pion and kaon decay constants from the NLO and NNLO fits.
 The left-most points for $F_\pi$ and $F_K$ correspond to the physical  experimental inputs.
The remaining lattice simulation data are taken from RBC and UKQCD~\cite{rbcukqcd11prd,rbcukqcd13prd}, where we have only included
the unitary points simulated with the physical strange quark mass.
The shaded area around each curve stands for the statistical error band from the fits. 
The meaning of notations for different lines are explained in detail in the text. 
}\label{figfk}
 \end{center}
\end{figure}

\begin{figure}[ht]
\begin{center}
 \includegraphics[angle=0.0,width=1.0\textwidth]{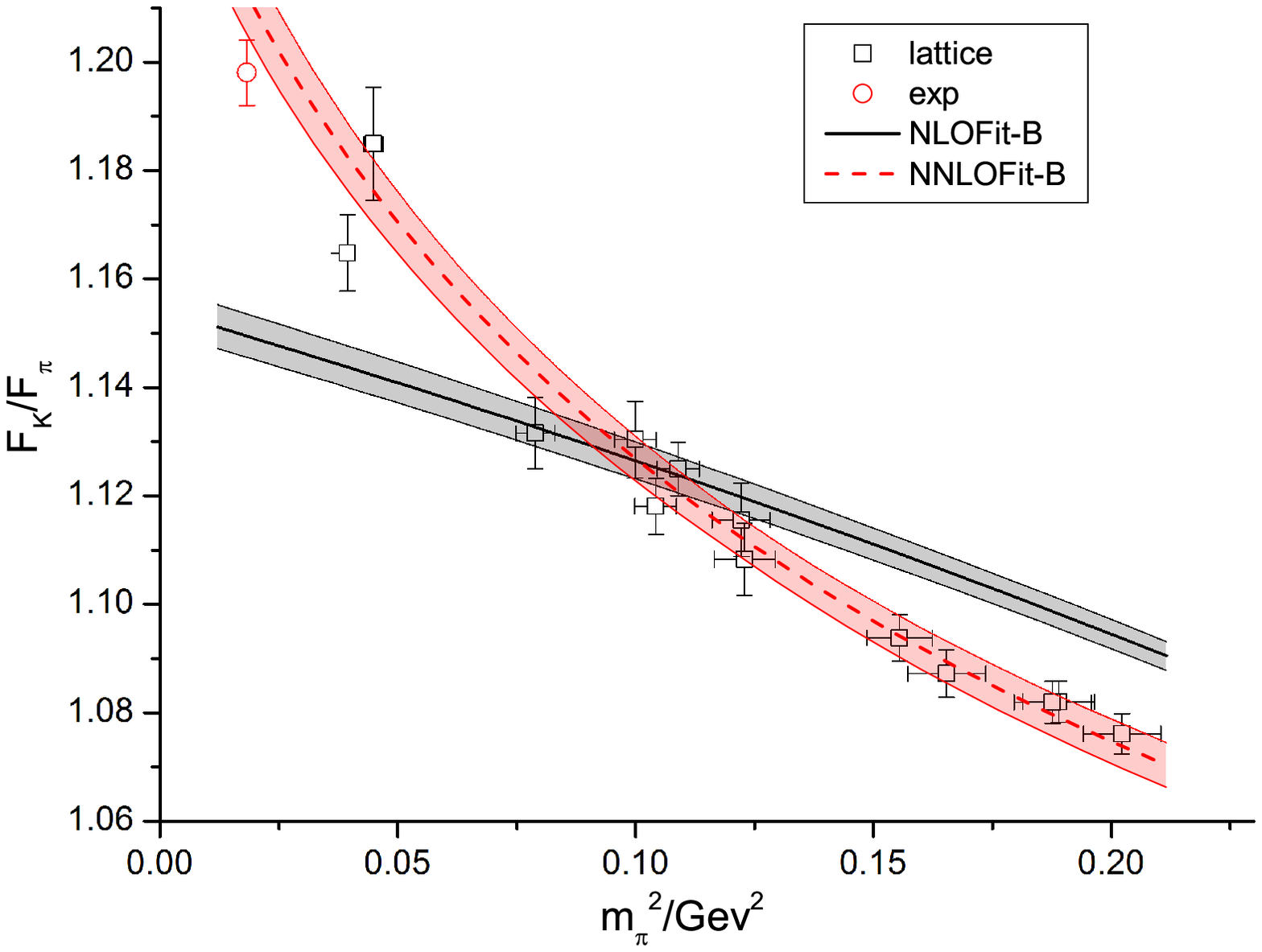}
\caption{  Ratio $F_K/F_\pi$  from the NLO and NNLO fits. The left most point corresponds to the experimental input.
The remaining lattice simulation data are taken from Ref.~\cite{Durr:2010hr} (BMW).
The shaded area around each curve stands for the statistical error band from the fits. 
The meaning of notations for different lines are explained in detail in the text. 
}\label{figfr}
 \end{center}
\end{figure}

The other quantities in the fits are presented in Figs.~\ref{figmeta}, \ref{figmk}, \ref{figfk} and \ref{figfr}, together with
the lattice simulation data and the experimental inputs. We find that the final outputs from
NLOFit-A and NLOFit-B are quite similar, so only the
plots from NLOFit-B are given explicitly.
The shaded area surrounding each curve corresponds to the statistical error band for each quantity.
In Fig.~\ref{figmeta}, we show the resulting figures from NLOFit-B for the masses of $\eta$ and $\eta'$.
In Figs.~\ref{figmk},~\ref{figfk} and \ref{figfr}, we show the corresponding plots for $m_K^2$, $F_{\pi,K}$
and $F_K/F_\pi$ as functions of $m_\pi^2$, respectively.

\begin{table}[H]
\centering
\begin{tabular}{c|c|c  }
\hline
                                & \makecell{NLOFit-A }               &\makecell{NLOFit-B  }            \\
 \hline
$\chi^2/(d.o.f)$                & \makecell{ 481.2/(76-5)}         &\makecell{477.7/(76-6)}       \\
\hline
$ M_{0} $~(MeV)                 & \makecell{835.7*}                 & \makecell{767.3$\pm$31.5$\pm$32.3}       \\
\hline
$ F $~(MeV)                     & \makecell{92.1$ \pm $0.2$\pm$0.6}   & \makecell{92.1$ \pm $0.2$\pm$0.6}    \\
\hline
$10^3\times L_{5} $           & \makecell{1.45$ \pm $0.02$\pm$0.30}  & \makecell{1.47$ \pm $0.02$\pm$0.29}      \\
\hline
$10^3\times  L_{8} $         & \makecell{1.00$ \pm $0.07$\pm$0.10}  & \makecell{1.08$ \pm $0.05$\pm$0.04}   \\
\hline
$ \Lambda_{1} $                 &\makecell{0.02$ \pm $0.05$\pm$0.06}    & \makecell{-0.09$ \pm $0.08$\pm$0.02}    \\
\hline
$\Lambda_{2} $                   &\makecell{0.25$ \pm $0.06$\pm$0.02}  &\makecell{0.14$ \pm $0.07$\pm$0.03}      \\
\hline
\end{tabular}
\caption{Parameters from the NLO fits. The meaning of different notations to label different fits are explained in detail in the text. In the
row of $M_0$, the columns with 835.7* denote the fit results by fixing the value of $M_0$ from its LO determination.
The first error bar for each parameter is the statistical one given by the fits and the second one corresponds to the systematic error.
The way to estimate the systematic error is explained in detail in the text.  }\label{tabnlofit}
\end{table}

\begin{table}[H]
\centering
\begin{tabular}{c|c|c|c }
\hline
  {Parameters}         & \makecell{Inputs}             & \makecell{NLOFit-A}                 & \makecell{NLOFit-B}        \\
\hline
$ F_0 $~(MeV)             &\makecell{118.0 $ \pm $16.5} &\makecell{104.9$\pm$2.9$\pm$0.3}    & \makecell{99.7$\pm$3.6$\pm$1.6}       \\
\hline
$ F_8 $~(MeV)              &\makecell{133.7 $ \pm $11.1}  & \makecell{113.2$\pm$0.3$\pm$4.4}   & \makecell{ 113.5$\pm$0.3$\pm$4.2}    \\
\hline
$ \theta_{0} $~(Degree)    &\makecell{-11.0 $ \pm $3.0}   &\makecell{-7.2$\pm$2.1$\pm$1.3}    & \makecell{ -10.6$\pm$2.4$\pm$0.1}      \\
\hline
$ \theta_{8} $~(Degree)   &\makecell{-26.7 $ \pm $5.4}    & \makecell{-21.5$\pm$2.2$\pm$3.9}   & \makecell{-25.4$\pm$2.6$\pm$2.3}  \\
\hline
$ m_s/\widehat{m} $       &\makecell{27.5 $ \pm $3.0}    &\makecell{22.6$\pm$0.8$\pm$0.6}     & \makecell{21.9$\pm$0.6$\pm$1.2}        \\
\hline
$ F_q $~(MeV)      &\makecell{$106.0\pm 11.1$*}           & \makecell{94.1$\pm$1.9$\pm$1.7}   & \makecell{ 90.6$\pm$2.4$\pm$0.4}      \\
\hline
$ F_s $~(MeV)       &\makecell{$143.8\pm16.5$*}            &\makecell{122.3$\pm$1.2$\pm$5.1}   & \makecell{120.9$\pm$1.2$\pm$5.5}     \\
\hline
$ \theta_{q} $~(Degree) &\makecell{$34.5\pm 5.4$*}            & \makecell{40.4$\pm$3.1$\pm$3.6}   & \makecell{35.0$\pm$3.7$\pm$1.6}        \\
\hline
$ \theta_{s} $~(Degree &\makecell{$36.0\pm 4.2$*}             &\makecell{39.9$\pm$1.7$\pm$2.2}     & \makecell{37.2$\pm$1.8$\pm$1.1}      \\
\hline
\end{tabular}
\caption{The outputs from NLO fits. Notice that $F_q, F_s, \theta_q$ and $\theta_s$ are not the phenomenological inputs in the fits, since
 they are related to $F_0, F_8, \theta_0$ and $\theta_8$ through Eq.~\eqref{fqfstof0f8}. The phenomenological values for the mixing parameters
are taken from Ref.~\cite{Chen:2014yta} and we triple the error bands here in order to make a conservative estimate. The input of $m_s/\widehat{m}$
is taken from the FLAG working group in Ref.~\cite{Aoki:2013ldr} and we assign the $10\%$ error bar as done in Ref.~\cite{Bijnens:2011tb}.
For the error bars of each quantity, the first one corresponds to the statistic error and the second one is for the systematic error, which
are explained in detail in the text. }\label{tabnloresult}
\end{table}

\subsection{  NLO fits focusing on the masses}

In this section, we present another kind of NLO fits by focusing on the masses of $\eta, \eta',K$ and
excluding the decay constants $F_\pi$, $F_K$ and their ratio.
This kind of discussion is well motivated, since it is known that the NNLO corrections in $\delta$ counting,
such as the LEC $ L_4$, are important to simultaneously describe $F_\pi$ and $F_K$~\cite{Ecker:2013pba,Bijnens:2011tb,Bijnens:2014lea}.
But this LEC is absent in NLO study.
We have also provided another independent confirmation on this finding in Fig.~\ref{figfk}, where one can see that the decay
constants of pion and kaon are poorly reproduced at next-to-leading order in $\delta$ expansion.
When only focusing on the $\eta$, $\eta'$ and $K$ masses and the ratio $m_s/\hat{m}$
at next-to-leading order the parameter $F$ can not be resolved, because it always appears in the form
$ L_5/F^2$ or $ L_8/F^2$. We will fix its value to $F=90$~MeV, close to the values given in Table~\ref{tabnlofit}.
For the mixing parameters we consider the mixing angles of $\theta_0$ and $\theta_8$, but exclude the constants $F_0$ and $F_8$.
This is because $F_0$ and $F_8$ are dependent on the parameter $F$ and should be determined together with $F_\pi$ and $F_K$.
For simplicity in later discussion, we call the fits performed in this section
as the mass-focusing type throughout.

As in the previous section, we present the fits with $F$ in the denominators of the theoretical expressions (e.g. Eq.~\eqref{fpinlof0})
and use the fits with $F_\pi$ to estimate the systematic errors.
For each case, we perform the fits either by fixing $M_0$ at its LO determination (NLOFit-C)
or by freeing its value (NLOFit-D). The fitted parameters are given in Table~\ref{tabnlomassfit}
and the $m_s/\hat{m}$ ratio and mixing angles are given in Table~\ref{tabnlomassresult}.
The resulting figures from NLOFit-C and NLOFit-D are quite similar and we explicitly show one set of them, e.g. NLOFit-D
in Figs.~\ref{figmeta} and \ref{figmk} for the $\eta^{(')}$ and kaon masses, respectively.

A significant difference between the results in Table~\ref{tabnlofit} and the mass-focusing fits
in Table~\ref{tabnlomassfit}
is that much larger statistical error bars are obtained in the latter case, especially for the LECs
$L_5$ and $L_8$, as they are constrained by fewer data.
Likewise, there are large systematic errors for the values of $L_5$ and $L_8$ in Table~\ref{tabnlomassfit},
indicating a larger truncation uncertainty due to higher orders.
%
%%%. Obvious changes also happen for the values of $ L_5$ and $ L_8$
%%%in the Fit2 case, comparing with the results in Table~\ref{tabnlofit} and Table~\ref{tabnlomassfit}.
%
We do not see a significant improvement when freeing the value of $M_0$ in the fits.

\begin{table}[H]
\centering
\begin{tabular}{c|c|c   }
\hline
                     & \makecell{NLOFit-C }           & \makecell{NLOFit-D }              \\
 \hline
$\chi^2/(d.o.f)$     & \makecell{ 168.8/(44-4)}     & \makecell{ 168.7/(44-5)}       \\
 \hline
$ M_{0} $~(MeV)      & \makecell{835.7*}             & \makecell{821.5$\pm$43.5}        \\
\hline
$10^3\times L_{5}$   & \makecell{1.40$ \pm $0.58$\pm$0.75}   & \makecell{1.51$ \pm $0.68$\pm$0.91}     \\
\hline
$10^3\times L_{8} $  & \makecell{0.88$ \pm $0.29$\pm$0.35}   & \makecell{0.94$ \pm $0.34$\pm$0.44}   \\
\hline
$ \Lambda_{1} $      &\makecell{-0.06$ \pm $0.04$\pm$0.02}   &\makecell{-0.09$ \pm $0.11$\pm$0.09}     \\
\hline
$\Lambda_{2} $       &\makecell{0.17$ \pm $0.19$\pm$0.25}    &\makecell{0.18$ \pm $0.19$\pm$0.25}     \\
\hline
\end{tabular}
\caption{ Parameters from the mass-focusing NLO fits. The meaning of different notations to label different fits are explained in detail in the text.
 $F$ is fixed at 90~MeV in these fits. The first error for each parameter corresponds to the statistical one and the second error denotes the systematic uncertainty.
 See the text for details.
 }\label{tabnlomassfit}
\end{table}

\begin{table}[H]
\centering
\begin{tabular}{c|c|c|c   }
\hline
  {Parameters}         & \makecell{Inputs}             & \makecell{NLOFit-C}                   & \makecell{NLOFit-D}        \\
\hline
$ \theta_{0} $~(Degree)    &\makecell{-11.0 $ \pm $3.0}   &\makecell{-10.6$\pm$2.4$\pm$3.3}    & \makecell{ -11.0$\pm$3.4$\pm$2.1}    \\
\hline
$ \theta_{8} $~(Degree)   &\makecell{-26.7 $ \pm $5.4}    & \makecell{-25.3$\pm$2.4$\pm$4.4}   & \makecell{-26.7$\pm$4.5$\pm$7.1}      \\
\hline
$ m_s/\widehat{m} $       &\makecell{27.5 $ \pm $3.0}    &\makecell{23.7$\pm$0.3$\pm$0.3}     & \makecell{23.6$\pm$0.5$\pm$0.1}          \\
\hline
$ \theta_{q} $~(Degree) &\makecell{$34.5\pm 5.4$*}            & \makecell{35.6$\pm$1.4$\pm$1.1}   & \makecell{34.1$\pm$4.3$\pm$4.1}         \\
\hline
$ \theta_{s} $~(Degree)  &\makecell{$36.0\pm 4.2$*}             &\makecell{37.0$\pm$0.9$\pm$0.7}   & \makecell{ 36.3$\pm$2.2$\pm$2.1}     \\
\hline
\end{tabular}
\caption{ The outputs from the mass-focusing NLO fits. See Table~\ref{tabnloresult} for the phenomenological inputs.
The first error for each quantity corresponds to the statistical one and the second error denotes the systematic uncertainty.
 See the text for details. }\label{tabnlomassresult}
\end{table}

\subsection{ Next-to-next-to-leading order analyses}\label{sectnnlo}

From the NLO discussions in the previous two sections, we observe that the phenomenological results and the lattice simulations on $\eta$ and $\eta'$ states
can be reasonably reproduced. This is an important improvement comparing with the LO study, since at this order we only have the conventional
one-mixing-angle formalism.
  The two-mixing-angle formalism only shows up beyond LO.
However, observing $m_K$, $F_\pi$, $F_K$ and their ratio in
Figs.~\ref{figmk},~\ref{figfk} and~\ref{figfr},
 it is clear that the  NLO analysis is still inadequate. We
need to include higher order contributions beyond NLO in order to further improve the descriptions.
Moreover, the chiral logarithms predicted by $\chi$PT at one loop start at NNLO in the $\delta$ expansion. Due to their
importance in other observables, we consider it is relevant to discuss the impact of these chiral logs.

As in the NLO case,   we perform two types of fits, using the NLO and NNLO theoretical expressions given in terms of
$F$ and $F_\pi$ for various observables.
We explicitly present the fit results with $F$ in the
theoretical expressions and use the alternative fits expressed in terms of $F_\pi$
to estimate the systematic errors,  due to working up to NNLO in $\delta$ and neglecting higher orders.
According to the Lagrangian in Eq.~\eqref{lagnnlo}, eleven additional unknown LECs appear at NNLO
and there will be seventeen parameters in total for the NNLO study.
At the present precision of the lattice simulations and phenomenological inputs, it is impossible to obtain sensible and stable fits if we free
all of the seventeen parameters. Therefore, we need to take other independent determinations for some of the LECs in order to proceed the NNLO study.

We mention that the state-of-art determinations of the $\cO(p^4)$ LECs in $SU(3)$ $\chi$PT suffer
 uncertainties from the many poorly
known $\cO(p^6)$ LECs~\cite{Bijnens:2014lea,Bijnens:2011tb}. Because of the large number of barely known $\cO(p^6)$ LECs, it is rather difficult to get conclusive results
in the present two-loop $SU(3)$ $\chi$PT studies~\cite{Bijnens:2014lea,Bijnens:2011tb}.
In the present work, there are five $\cO(p^6)$ LECs, i.e. $C_{12},C_{14},C_{17},C_{19},C_{31}$, in Eq.~\eqref{lagnnlo}
and we cannot make precise determinations of these $C_i$ parameters here. Maybe when taking into account the scattering data,
one can make more stringent constraints on the $C_i$ LECs in $U(3)$ $\chi$PT. But this is beyond the scope of current work.
Instead we take the $C_i$ values from the Dyson-Schwinger--like approach given in Ref.~\cite{Jiang:2009uf}, where all of the $\cO(p^6)$
$C_i$ at leading $N_C$ are predicted. In order to show the dependences of the final results on the $C_i$ values, we also perform
other fits by using their updated determinations~\cite{Jiang15}.
Like in Ref.~\cite{Bijnens:2011tb}, we multiply the $\cO(p^6)$ $C_i$ from Refs.~\cite{Jiang:2009uf,Jiang15}
by a global factor $\alpha$ and consider $\alpha$ as a free parameter in the fits. In this way, we partially compensate the large uncertainties
of the $C_i$ parameters.

For the operators proportional to $v_2^{(2)}$, $ L_{18}$ and $ L_{25}$ in Eq.~\eqref{lagnnlo}, they are not present in $SU(3)$ $\chi$PT and purely contribute to
the $\eta$-$\eta'$ mixing, being irrelevant to the pion and kaon observables. Since the $\eta$-$\eta'$ mixing parameters have already been
satisfactorily described in the NLO fits, we do not further include $v_2^{(2)}$, $ L_{18}$ and $ L_{25}$ at 
NNLO study. \footnote{ Indeed, in this work $M_0$ and $v_2^{(2)}$ only enter in the mass Lagrangian in Eq.~\eqref{lagmixingpara2} explicitly. 
They always appear combined in the effective form $M_{0,\, eff}^2=M_0^2 + 6 v_2^{(2)} ( 2m_K^2+m_\pi^2)$, which is the parameter we are actually extracting. 
The contribution $v_2^{(2)}$ could be singled out through the study of the $\eta_0\eta_0\to \pi\pi$ scattering. 
However we point out that the anti-correlation between $M_0$ and $v_2^{(2)}$ in general can not be 
recovered in the present numerical fits, due to the presence of far too many parameters in the problem and the large uncertainties of the lattice simulation data, specially 
for the determinations of $m_{\eta'}$.  }
Their inclusion in the present analysis tend to make the fit unstable.
Clearly studying more $\eta^{(')}$ related observables it would be possible to extract these parameters but this
is out of the reach of the present analysis. A global fit is too unconstrained, being unstable and producing
values of the latter couplings compatible with zero within uncertainties.
Then we are left with three $\cO(N_C^0,p^4)$ operators: $ L_4$, $ L_6$ and $ L_7$, which have corresponding parts in $SU(3)$ $\chi$PT.
Since $U(3)$ and $SU(3)$ $\chi$PT contain different dynamical degrees of freedom, the corresponding LECs from the two theories can be different.
A typical example is the $L_7$ parameter in $SU(3)$ $\chi$PT, which is demonstrated to be dominated by the singlet $\eta_0$ state~\cite{gasser8485}.
Since in $U(3)$ $\chi$PT the singlet $\eta_0$ has been explicitly introduced, the value of $ L_7$ in this theory can be
totally different from $L_7^{SU(3)}$ in $SU(3)$ case. While for other $\cO(p^4)$ LEcs, such as $ L_{i=4,5,6,8}$, the differences between $U(3)$
and $SU(3)$ $\chi$PT are not expected to be as large as the $ L_7$ case, since they do not receive the tree-level contributions
from the $\eta_0$ state.

Another subtlety to take into account is that $m_\eta$ and $m_{\eta'}$ appear in the chiral loops and, at the same time,
the final expressions of $m_\eta$ and $m_{\eta'}$
depend on the these loops as well. In order to avoid making the complicated iterative procedure to obtain the $\eta$-$\eta'$ mixing parameters, we
use the LO formulas for $m_\eta$ and $m_{\eta'}$ in the chiral loops. The differences caused by this simple treatment and the strict
iterative procedure are beyond the NNLO precision in $\delta$ expansion, since the chiral loops themselves
are already NNLO.
Our simple solution is also justified by the fact that the LO description of $m_\eta$ and $m_{\eta'}$ is in qualitative agreement with
the lattice simulation data, as shown in Sect.~\ref{sectlo}. Since the qualitative agreement between the LO formulas and the lattice simulation data
requires the value of $M_0$ to be around 835.7~MeV, as given in Eq.~\eqref{lofit}, we fix $M_0=835.7$~MeV in the following discussions.
 This also helps to stabilize the NNLO fits, with its many free parameters.
Other useful criteria to discriminate reasonable fits are the a priori ranges
estimated in Eq.~\eqref{lam12range}, since the fits with large magnitudes of $\Lambda_1$ and $\Lambda_2$ imply unphysically large corrections to
the $\eta$-$\eta'$ mixing parameters
 and the breakdown of the $\delta$ expansion.
In the following we only present the fit results that are consistent with Eq.~\eqref{lam12range}.

With all of the above setups, the values of parameters from the NNLO fits are summarized in Table~\ref{tabnnlofit}.
The fits labeled by NNLOFit-A and NNLOFit-B  correspond to using different values of the $\cO(p^6)$ LECs.
For NNLOFit-A, the $C_i$ values are taken from Ref.~\cite{Jiang:2009uf}:
\begin{eqnarray}\label{ci10}
 C_{12}=-0.34\,,\quad C_{14}=-0.83\,,\quad C_{17}= 0.01\,, \quad C_{19}=-0.48\,, \quad C_{31}=-0.63\,,
\end{eqnarray}
which are given in units of $10^{-3}$GeV$^{-2}$.
For NNLOFit-B, we take   their updated   $\cO(p^6)$ $C_i$ values from Ref.~\cite{Jiang15}:
\begin{eqnarray}\label{ci15}
 C_{12}=-0.34\,,\quad C_{14}=-0.87\,,\quad C_{17}=0.17\,, \quad C_{19}=-0.27\,, \quad C_{31}=-0.46\,,
 \end{eqnarray}
in the same units as before.

It is clear that the parameters resulting from
 fits with different $C_i$ inputs slightly differ from one another.
We remind that the first error bar for each parameter in Table~\ref{tabnnlofit} corresponds to the
statistical one directly from the fits and the second error bar stands for the systematic one, which
is estimated, as usual, from the variation of the parameter from the NNLO fits with the theoretical
expressions in terms of $F$ and those expressed as functions of $F_\pi$.

At NNLO, one has the contributions from the chiral loops and
the $\cO(p^6)$ LECs, which make our determinations in Table~\ref{tabnnlofit} closer to the recent two-loop results of the
$SU(3)$ $\chi$PT LECs, comparing with the NLO determinations in Table~\ref{tabnlofit}.
Some typical trends of the values of parameters from the NLO study in Table~\ref{tabnlofit} to the NNLO one in Table~\ref{tabnnlofit}
 are summarized now.
The axial-vector decay constant $F$ at leading $N_C$ and chiral limit is reduced at NNLO, which is mainly due to the inclusion of $ L_4$.
Our conclusion is based on the fact that strong correlations between $F$ and $ L_4$ always appear, which has been confirmed in previous
study~\cite{Guo:2014yva,Ecker:2013pba}. For $ L_5$ and $ L_8$, we find that their values are obviously reduced  compared to   the NLO
determination and become closer to the two-loop results in Ref.~\cite{Bijnens:2014lea}. As mentioned in the former reference,
the discussions in the two-loop $SU(3)$ $\chi$PT are sensitive to the value of the $1/N_C$ suppressed LEC $ L_4$.
The present study provides an independent determination for this parameter and for the $1/N_C$ suppressed LEC $ L_6$ as well.
We mention that our determinations of $ L_4$ have opposite signs   with respect to that
in Ref.~\cite{Bijnens:2014lea}, which may be the source of the smaller
$F$ obtained in that reference. Notice that the present values of $L_4, L_5, L_6, L_8$ are rather compatible with the 
combinations of $2L_8-L_5$ and $2L_6-L_4$ given in Ref.~\cite{Oller:2006xb}. 
Fit solutions with larger $\Lambda_1$ and $\Lambda_2$ than those in Eq.~\eqref{lam12range} (out of the a priori range ~\eqref{lam12range})
are discarded: they are not considered as reasonable physical solutions and will not be discussed any further.
According to the values of $\alpha$ in the two fits,
it seems that our study somewhat prefers smaller magnitudes of the $\cO(p^6)$ LECs than those from
the Dyson-Schwinger approach given in Refs.~\cite{Jiang:2009uf,Jiang15} and also prefers a global change of sign with respect to
Eqs.~\eqref{ci10} and \eqref{ci15}. We have investigated the impact of fitting $\alpha$ but releasing one of the
$\cO(p^6)$ LECs as an independent parameter (e.g., $C_{14}$), but no definitive conclusion could be extracted.
These puzzles cannot be resolved here and it is definitely interesting and necessary to further investigate
the values of the $\cO(p^6)$ LECs in the future.

The various plots from the NNLO fits are shown in Fig.~\ref{figmeta} for $m_\eta$ and $m_{\eta'}$, Fig.~\ref{figmk} for $m_K$,
Fig.~\ref{figfk} for $F_\pi$ and $F_K$, and Fig.~\ref{figfr} for the ratio $F_K/F_\pi$, together with the NLO results and the lattice simulation data and
experimental inputs. The shaded area surrounding each curve represents the statistical error band. 
The figures from NNLOFit-B are compatible with those from NNLOFit-A within the uncertainties, so we only show the results 
for the former in Figs.~\ref{figmeta}, \ref{figmk}, \ref{figfk} and \ref{figfr}.

In addition, to demonstrate the effects by using the LO masses in the higher order corrections, instead of the renormalized ones, we 
explicitly show the results for $m_\eta$ and $m_{\eta'}$ expressed in terms of the LO masses $\overline{m}_\pi, \overline{m}_K$ and $1/F^2$ 
in Fig.~\ref{figmeta}, with the lines labeled as NNLOFit-B-$\overline{m}_{\pi, K}$. 
The values of the LECs when plotting these lines are exactly the same as those from 
the NNLOFit-B column in Table~\ref{tabnnlofit}.  In this way, one can directly see  
the differences due to the N$^3$LO truncation uncertainty caused by using the renormalized masses and the LO ones at the NNLO level.  
According to Fig.~\ref{figmeta}, we conclude that the differences for $m_\eta$ and $m_{\eta'}$ caused by using different types of masses 
in the higher order corrections are rather within the statistical uncertainties from the fits and therefore the differences 
should be perfectly compatible within the total uncertainties 
after taking into account the systematic ones in Table~\ref{tabnnlofit}. We verify that similar conclusions are obtained for other cases.
In order not to overload the plots in other figures, we shall not explicitly show the results given in 
terms of $\overline{m}_\pi$ and $\overline{m}_K$.

From Figs.~\ref{figmeta}, \ref{figmk}, \ref{figfk} and \ref{figfr}, we observe, when compared with the curves of the NLO study,  
slight improvements in the reproduction of the masses for $\eta,\eta'$ and significant ones for $m_K$, $F_\pi$, $F_K$ and the ratios
of $F_K/F_\pi$. 
Moreover the $\chi^2$ for the NNLO fits
are greatly reduced compared with $\chi^2$ for the NLO ones, indicating that
the NNLO corrections are important at the present level of precision
and essential to simultaneously describe the lattice
simulation data and experimental inputs of the light pseudoscalar mesons $\pi, K, \eta$ and $\eta'$.

\begin{table}[H]
\centering
\begin{tabular}{c|c|c  }
\hline
                       & \makecell{NNLOFit-A }                  &\makecell{NNLOFit-B  }                   \\
 \hline
$\chi^2/(d.o.f)$       &\makecell{212.4/(76-9)}              & \makecell{ 231.9/(76-9)}             \\
\hline
$ F $~(MeV)            & \makecell{81.7$ \pm $1.5$\pm$5.3}   & \makecell{80.8$ \pm $1.6$\pm$6.1}     \\
\hline
$  10^3\times L_{5} $  & \makecell{0.60$ \pm $0.11$\pm$0.52}  & \makecell{0.45$ \pm $0.12$\pm$0.78}     \\
\hline
$  10^3\times L_{8} $  & \makecell{0.25$ \pm $0.07$\pm$0.31}   & \makecell{0.30$ \pm $0.06$\pm$0.30}   \\
\hline
$ \Lambda_{1} $        & \makecell{-0.003$ \pm $0.060$\pm$0.093} &\makecell{-0.04$ \pm $0.06$\pm$0.13}  \\
\hline
$\Lambda_{2} $        &\makecell{0.08$ \pm $0.11$\pm$0.20}  &\makecell{0.14$ \pm $0.10$\pm$0.40}  \\
\hline
$  10^3\times L_{4} $   & \makecell{-0.12$ \pm $0.06$\pm$0.19} &\makecell{-0.09$ \pm $0.06$\pm$0.23}   \\
\hline
$ 10^3\times L_{6} $  & \makecell{-0.05$ \pm $0.04$\pm$0.02}  & \makecell{0.03$ \pm $0.03$\pm$0.02}   \\
\hline
$ 10^3\times  L_{7} $   & \makecell{0.26$ \pm $0.05$\pm$0.06} & \makecell{0.36$ \pm $0.05$\pm$0.12}    \\
\hline
$\alpha $              &\makecell{-0.59$ \pm $0.09$\pm$0.18} & \makecell{-0.76$ \pm $0.08$\pm$0.44}     \\
\hline
\end{tabular}
\caption{Parameters from the NNLO fits. In all of these fits, $M_0$ is fixed at 835.7~MeV from its LO determination.
The meaning of different notations to label different fits are explained in detail in the text.
The first error bar for each parameter corresponds to the statistical one and the second error denotes the systematic uncertainty.
 See the text for details.  }\label{tabnnlofit}
\end{table}

\begin{table}[H]
\centering
\begin{tabular}{c|c|c|c }
\hline
  {Parameters}            & \makecell{Inputs}           & \makecell{NNLOFit-A}    & \makecell{NNLOFit-B}      \\
\hline
$ F_0 $~(MeV)            &\makecell{118.0 $ \pm $16.5} & \makecell{108.0$\pm$1.5$\pm$3.6}   &\makecell{109.1$\pm$1.3$\pm$5.9}     \\
\hline
$ F_8 $~(MeV)            &\makecell{133.7 $ \pm $11.1} & \makecell{ 124.7$\pm$1.2$\pm$8.7}   &\makecell{126.5$\pm$1.2$\pm$11.8}    \\
\hline
$ \theta_{0} $~(Degree)  &\makecell{-11.0 $ \pm $3.0}  & \makecell{ -6.8$\pm$1.1$\pm$2.6}     &\makecell{-6.8$\pm$0.9$\pm$3.7}    \\
\hline
$ \theta_{8} $~(Degree)  &\makecell{-26.7 $ \pm $5.4}   & \makecell{-26.8$\pm$1.1$\pm$0.2}  & \makecell{-27.9$\pm$1.0$\pm$1.4}  \\
\hline
$ m_s/\widehat{m} $      &\makecell{27.5 $ \pm $3.0}   & \makecell{27.0$\pm$0.6$\pm$0.4}    &\makecell{29.4$\pm$0.4$\pm$0.6}       \\
\hline
$ F_q $~(MeV)            &\makecell{$106.0\pm 11.1$*} & \makecell{ 92.8$\pm$1.1$\pm$1.2}    & \makecell{92.7$\pm$1.0$\pm$1.0}    \\
\hline
$ F_s $~(MeV)            &\makecell{$143.8\pm16.5$*}  & \makecell{136.4$\pm$1.5$\pm$10.0}     &\makecell{139.0$\pm$1.4$\pm$14.9}     \\
\hline
$ \theta_{q} $~(Degree)  &\makecell{$34.5\pm 5.4$*}     & \makecell{36.4$\pm$1.4$\pm$0.2}   &\makecell{35.8$\pm$1.2$\pm$0.3}      \\
\hline
$ \theta_{s} $~(Degree)   &\makecell{$36.0\pm 4.2$*}     & \makecell{37.8$\pm$0.9$\pm$1.5}    &\makecell{37.1$\pm$0.8$\pm$1.1}     \\
\hline
\end{tabular}
\caption{The outputs from NNLO fits. See Table~\ref{tabnloresult} for the explanation of the phenomenological inputs.
The first error for each quantity corresponds to the statistical one and the second error denotes the systematic one.
 See the text for details.  }\label{tabnnloresult}
\end{table}

\section{Conclusions}\label{sec.concl}

  In this article we have performed
a thorough study on the $\eta$-$\eta'$ mixing, and axial-vector decay constants for the pion and kaon,
up to next-to-next-to-leading order in $\delta$ expansion within
  $U(3)$ chiral perturbation theory.
We have carried on a detailed scrutiny  and discussions of our results, which have been carefully compared to
other works in literature for the $\eta$-$\eta'$ mixing.
A general mixing formalism, including the higher-derivative terms and kinematic mixing cases, has been addressed in detail.
The connections between the mixing parameters from the popular two-mixing-angle scheme and the low energy constants from chiral
perturbation theory  have been established, both for the singlet-octet basis and the quark-flavor basis.

The considered quantities, including the masses of $\eta,\eta'$ and $K$, the quark mass ratio of $m_s/\widehat{m}$,
the parameters in the two-mixing-angle scheme
and the $\pi, K$ decay constants have been confronted with recent lattice simulations and phenomenological inputs.
 We find that the next-to-leading-order fits yield satisfactory descriptions
for the masses of the three pseudoscalar mesons as functions of $m_\pi^2$ and the four mixing parameters ($F_0,F_8,\theta_0,\theta_8$),
producing in addition reasonable values of low energy constants.
Nonetheless,  when the $\pi$ and $K$ decay constants are included together with the masses and mixing parameters in the fits,
the next-to-leading-order analyses are inadequate and
it is necessary to step into the next-to-next-to-leading-order study.
 Using the $\cO(p^6)$ LECs determinations from a Dyson-Schwinger-like approach~\cite{Jiang:2009uf,Jiang15} multiplied by a global factor,
we are able to achieve a reasonable description for all of the physical quantities considered above and the resulting values for
the leading $N_C$ $\cO(p^4)$ low energy constants $L_5$ and $L_8$ turn to be compatible with the very recent two-loop determinations
in Ref.~\cite{Bijnens:2014lea}. Therefore we conclude that the large $N_C$ $U(3)$ chiral perturbation theory offers a concise
theoretical framework that is able to simultaneously reproduce accurately the general $\eta$-$\eta'$ mixing
and to provide sophisticated enough expressions to describe
the chiral extrapolations of the $\pi$ and $K$ decay constants and masses.

Our results are also useful for future phenomenological studies of different processes involving $\eta$ and $\eta'$.  
Combining Eq.~\eqref{twoanglesmixing08} or Eq.~\eqref{twoanglesmixingqs} with Table~\ref{tabnnloresult}, one can directly find the 
relations between the physical states $\eta,\eta'$ and the octet-singlet bases $\eta_8, \eta_0$ or the quark-flavor bases $\eta_q,\eta_s$. 
These relations are consistent with the requirements from the recent lattice simulations and phenomenology. 

Finally, it is worthy to remark that some of the parameters in our best analysis (NNLOFit-B) in Table~\ref{tabnnlofit}
have been determined with relatively small errors.
For instance, the NLO parameters $\Lambda_{1,2}$, which are fitted up to $\cO(N_C^{-2})$ in the NNLO analysis, become
\begin{eqnarray}
\Lambda_{1} \, =\,-0.04\pm 0.06\pm0.13\, , \qquad \qquad  \Lambda_{2} \, =\,0.14 \pm 0.10\pm 0.40\, .
\end{eqnarray}
The NNLO fit also determines some $U(3)$ NNLO couplings with relatively high precision. NNLOFit-B yields
\begin{eqnarray}
10^3\times L_{4} \, =\, -0.09  \pm  0.06 \pm 0.23\, ,\qquad
10^3\times L_{6} \,=\, 0.03 \pm 0.03 \pm 0.02\, ,\qquad
10^3\times  L_{7} \, =\, 0.36  \pm  0.05 \pm 0.12\, .
\end{eqnarray}
Even though the error estimates in the present article must be considered with some caution,
as some lattice systematic uncertainties escape our control,
this hints the potentiality of this $U(3)$ $\chi$PT framework. We hope these results may encourage future lattice analyses along this line.

\section*{Acknowledgments}
We thank Shao-Zhou Jiang for communication on the updated values of the $\cO(p^6)$ LECs.
This work is supported in part by the National Natural Science Foundation of
China (NSFC) under Grant No. 11105038, the Natural Science Foundation of Hebei Province with contract No.~A2015205205, 
the grants from the Education Department of Hebei Province under contract No.~YQ2014034, 
the grants from the Department of Human Resources and Social Security of Hebei Province with contract No.~C201400323, 
and the Doctor Foundation of Hebei Normal University under Contract No.~L2010B04, 
  the Spanish Government (MINECO) and the European Commission (ERDF) [FPA2010-17747, FPA2013-44773-P, FPA2013-40483-P, 
SEV-2012-0249 (Severo Ochoa Program), CSD2007-00042 (Consolider Project CPAN)], the grants with contract No. FIS2014-57026-REDT from MINECO (Spain), 
and EPOS network of the European Community Research Infrastructure Integrating Activity ``Study of Strongly Interacting Matter'' (HadronPhysics3, Grant No. 283286). 
J.J. Sanz-Cillero wants to thank the Center for Future High Energy Physics and the Institute of High Energy Physics in Beijing for their hospitality. 

\appendix

\section{Higher order corrections to the $\overline{\eta}$ and $\overline{\eta}'$ bilinear terms}\label{app.delta}

In the following we provide the explicit expressions of the $\delta_i$'s in Eq.~\eqref{lagmixingpara}.
When expressing the results in terms of $F$, they take the form
\begin{eqnarray}
\delta_1= \dfrac{32C_{12}}{3F^2}\big[ c_{\theta}^{2}(4m_{K}^{2}-m_{\pi}^{2})
+4\sqrt{2}c_{\theta}s_{\theta}(m_{K}^{2}-m_{\pi}^{2})+s_{\theta}^{2}(2m_{K}^{2}+m_{\pi}^{2}) \big]\,,
\end{eqnarray}
\begin{eqnarray}
\delta_2= \dfrac{32C_{12}}{3F^2}\big[ c_{\theta}^{2} (2m_{K}^{2}+m_{\pi}^{2})
-4\sqrt{2}c_{\theta}s_{\theta}(m_{K}^{2}-m_{\pi}^{2})+s_{\theta}^{2}(4m_{K}^{2}-m_{\pi}^{2}) \big]\,,
\end{eqnarray}
\begin{eqnarray}
\delta_3= -\dfrac{64C_{12}}{3F^2}(m_{K}^{2}-m_{\pi}^{2}) \big( \sqrt2 c_{\theta}^{2} -c_{\theta}s_{\theta} - \sqrt2 s_{\theta}^2 \big)\,,
\end{eqnarray}
\begin{eqnarray}
 \delta_{\overline{\eta}}=&&\dfrac{8 L_{5}[c_{\theta}^{2}(4m_{K}^{2}-m_{\pi}^{2})+
4\sqrt{2}c_{\theta}(m_{K}^{2}-m_{\pi}^{2})s_{\theta}+(2m_{K}^{2}+m_{\pi}^{2})s_{\theta}^{2}]}{3F^{2}} +s_{\theta}^{2}\Lambda_{1}\nn \\ &&
+\dfrac{c_{\theta}^{2}A_{0}(m_{K}^{2})}{16\pi^{2}F^{2}} +\dfrac{8 L_{4}(2m_{K}^{2}+m_{\pi}^{2})}{F^{2}}+\dfrac{8 L_{18}s_{\theta}[2\sqrt{2}c_{\theta}
(m_{K}^{2}-m_{\pi}^{2})+(2m_{K}^{2}+m_{\pi}^{2})s_{\theta}]}{F^{2}}\nn \\
&& +\dfrac{64 L_{5}( L_{5}-2 L_{8})[ c_\theta^{2}(4m_{K}^{4}-m_{\pi}^{4})+4\sqrt{2}c_\theta(m_{K}^{4}-m_{\pi}^{4})s_\theta+(2m_{K}^{4}
+m_{\pi}^{4})s_\theta^{2} ] }{3F^{4}}\nn \\&&
+\dfrac{16(C_{14}+C_{17})}{3F^{2}}
[c_{\theta}^{2}(8m_{K}^{4}-8m_{K}^{2}m_{\pi}^{2}+3m_{\pi}^{4})+8\sqrt{2}c_{\theta}m_{K}^{2}(m_{K}^{2}-m_{\pi}^{2})s_\theta+
 (4m_{K}^{4}-4m_{K}^{2}
m_{\pi}^{2}+3m_{\pi}^{4})s_{\theta}^{2}]\,, \nn \\
\end{eqnarray}
\begin{eqnarray}
\delta_{\overline{\eta}'}=&& \dfrac{8 L_{5}[c_{\theta}^{2}(2m_{K}^{2}
+m_{\pi}^{2})+4\sqrt{2}c_{\theta}(-m_{K}^{2}+m_{\pi}^{2})s_{\theta}+(4m_{K}^{2}-m_{\pi}^{2})s_{\theta}^{2}]}{3F^{2}}+ c_{\theta}^{2}\Lambda_{1}
\nn \\&& + \dfrac{s_{\theta}^{2}A_{0}(m_{K}^{2})}{16\pi^{2}F^{2}}+\dfrac{8 L_{4}(2m_{K}^{2}+m_{\pi}^{2})}{F^{2}}+\dfrac{8 L_{18}c_{\theta}[c_{\theta}
(2m_{K}^{2}+m_{\pi}^{2})+2\sqrt{2}(-m_{K}^{2}+m_{\pi}^{2})s_{\theta}]}{F^{2}} \nn \\ &&
+\dfrac{64 L_{5}( L_{5}-2 L_{8})[c_\theta^{2}(2m_{K}^{4}+m_{\pi}^{4})+4\sqrt{2}c_\theta(-m_{K}^{4}+m_{\pi}^{4})s_\theta+(4m_{K}^{4}
-m_{\pi}^{4})s_\theta^{2}] }{3F^{4}}\nn \\&&
 +\dfrac{16(C_{14}+C_{17})}{3F^{2}}[c_{\theta}^{2}(4m_{K}^4-4m_{K}^{2}m_{\pi}^{2}+3m_{\pi}^{4})+8\sqrt{2}c_{\theta}m_{K}^{2}
(-m_{K}^{2}+m_{\pi}^{2})s_{\theta}+(8m_{K}^{4}-8m_{K}^{2}m_{\pi}^{2}+3m_{\pi}^{4})s_{\theta}^{2}]\,, \nn \\
\end{eqnarray}
\begin{eqnarray}\label{deltaknnlof0}
\delta_{k}= && -\dfrac{16 L_{5}(m_{K}^{2}-m_{\pi}^{2})
(\sqrt{2}c_{\theta}^{2}-c_{\theta}s_{\theta}-\sqrt{2}s_{\theta}^{2})}{3F^{2}}-c_{\theta}s_{\theta}\Lambda_{1}\nn \\&&
+\dfrac{c_{\theta}s_{\theta}A_{0}(m_{K}^{2})}{16\pi^{2}F^{2}}
-\dfrac{8 L_{18}[\sqrt{2}c_{\theta}^{2}(m_{K}^{2}-m_{\pi}^{2})+c_{\theta}(2m_{K}^{2}+m_{\pi}^{2})s_{\theta}+\sqrt{2}
(-m_{K}^{2}+m_{\pi}^{2})s_{\theta}^{2}]}{F^{2}} \nn \\ &&
 -\dfrac{128 L_{5}( L_{5}-2 L_{8})(m_{K}^{4}-m_{\pi}^{4})(\sqrt{2}c_\theta^{2}-c_\theta s_\theta-\sqrt{2}s_\theta^{2})}{3F^{4}}\nn \\ &&
 -\dfrac{64(C_{14}+C_{17})m_{K}^{2}(m_{K}^{2}-m_{\pi}^{2})(\sqrt{2}c_{\theta}^{2}-c_{\theta}s_{\theta}-\sqrt{2}s_{\theta}^{2})}
{3F^{2}} \,,
\end{eqnarray}
\begin{eqnarray}
 \delta_{m_{\overline{\eta}}^{2}}=&& \dfrac{16 L_{8}}{3F^{2}}[c_{\theta}^{2}(8m_{K}^{4}-8m_{K}^{2}m_{\pi}^{2}+3m_{\pi}^{4})+8\sqrt{2}c_{\theta}m_{K}^{2}
(m_{K}^{2}-m_{\pi}^{2})s_{\theta}+(4m_{K}^{4}-4m_{K}^{2}m_{\pi}^{2}+3m_{\pi}^{4})s_{\theta}^{2}]\nn \\ &&
 +\dfrac{2}{3}s_{\theta}[2\sqrt{2}c_{\theta}(m_{K}^{2}-m_{\pi}^{2})+(2m_{K}^{2}+m_{\pi}^{2})s_{\theta}]\Lambda_{2}
\nn \\&& + \dfrac{1}{16\pi^{2}}\bigg\{ \dfrac{1}{18F^{2}} \bigg[ c_{\theta}^{4}(16m_{K}^{2}-7m_{\pi}^{2})
+4\sqrt{2}c_{\theta}^{3}(8m_{K}^{2}-5m_{\pi}^{2})s_{\theta}+12c_{\theta}^{2}(4m_{K}^{2}-m_{\pi}^{2})s_{\theta}^{2}
\nn \\&&
 \qquad +16\sqrt{2}c_{\theta}(m_{K}^{2}-m_{\pi}^{2})s_{\theta}^{3}+2(2m_{K}^{2}+m_{\pi}^{2})s_{\theta}^{4} \bigg]A_{0}(m_{\eta}^{2}) \nn \\&&
 \qquad+\dfrac{(4m_{K}^{2}-m_{\pi}^{2})(2c_{\theta}^{4}-2\sqrt{2}c_{\theta}^{3}s_{\theta}-3c_{\theta}^{2}s_{\theta}^{2}+
2\sqrt{2}c_{\theta}s_{\theta}^{3}+2s_{\theta}^{4})}{18F^{2}}A_{0}(m_{\eta'}^2)\nn \\&&
\qquad -\dfrac{[c_{\theta}^{2}m_{\pi}^{2}+2\sqrt{2}c_{\theta}(-2m_{K}^{2}+m_{\pi}^{2})s_{\theta}-4m_{K}^{2}s_{\theta}^{2}]
}{3F^{2}}A_{0}(m_{K}^{2})+\dfrac{m_{\pi}^{2}(c_{\theta}^{2}-2\sqrt{2}c_{\theta}s_{\theta}+2s_{\theta}^{2})
}{2F^{2}}A_{0}(m_{\pi}^{2}) \bigg\} \nn \\&&
-\dfrac{16 L_{25}s_{\theta}[4\sqrt{2}c_{\theta}m_{K}^{2}(m_{K}^{2}-m_{\pi}^{2})+(4m_{K}^{4}-4m_{K}^{2}m_{\pi}^{2}
+3m_{\pi}^{4})s_{\theta}]}{F^{2}} +6(2m_{K}^{2}+m_{\pi}^{2})s_{\theta}^{2}v_2^{(2)}\nn \\&&
+\dfrac{16 L_{6}(2m_{K}^{2}+m_{\pi}^{2})[c_{\theta}^{2}(4m_{K}^{2}-m_{\pi}^{2})+4\sqrt{2}c_{\theta}(m_{K}^{2}-m_{\pi}^{2})
s_{\theta}+(2m_{K}^{2}+m_{\pi}^{2})s_{\theta}^{2}]}{3F^{2}}\nn \\&&
+\dfrac{16 L_{7}[8c_{\theta}^{2}(m_{K}^{2}-m_{\pi}^{2})^{2}+4\sqrt{2}c_{\theta}(2m_{K}^{4}-m_{K}^{2}m_{\pi}^{2}-m_{\pi}^{4})
s_{\theta}+(2m_{K}^{2}+m_{\pi}^{2})^{2}s_{\theta}^{2}]}{3F^{2}}\nn \\&&
+\dfrac{256( L_{5}-2 L_{8}) L_{8}}{3F^{4}} \bigg[c_\theta^{2}(8m_{K}^{6}-4m_{K}^{4}m_{\pi}^{2}-4m_{K}^{2}m_{\pi}^{4}+3m_{\pi}^{6})+\nn \\&&
\qquad 4\sqrt{2}c_\theta
m_{K}^{2}(2m_{K}^{4}-m_{K}^{2}m_{\pi}^{2}-m_{\pi}^{4})s_\theta+(4m_{K}^{6}-2m_{K}^{4}m_{\pi}^{2}-2m_{K}^{2}m_{\pi}^{4}+3m_{\pi}^{6})s_\theta^{2} \bigg]
\nn \\&&+\dfrac{16( L_{5}-2 L_{8})\Lambda_2 s_\theta[2\sqrt{2}c_\theta(m_{K}^{4}-m_{\pi}^{4})+(2m_{K}^{4}+m_{\pi}^{4})s_\theta]}{3F^{2}}\nn \\&&
+\dfrac{16(3C_{19}+2C_{31})}{3F^{2}} \bigg[c_{\theta}^{2}(16m_{K}^{6}-24m_{K}^{4}m_{\pi}^{2}+12m_{K}^{2}m_{\pi}^{4}-m_{\pi}^{6})
\nn \\&&
+4\sqrt{2}c_{\theta}(4m_{K}^{6}-6m_{K}^{4}m_{\pi}^{2}+3m_{K}^{2}m_{\pi}^{4}-m_{\pi}^{6})s_{\theta}
+(8m_{K}^{6}-12m_{K}^{4}m_{\pi}^{2}+6m_{K}^{2}m_{\pi}^{4}+m_{\pi}^{6})s_{\theta}^{2} \bigg]\,,
\end{eqnarray}
\begin{eqnarray}
\delta_{m_{\overline{\eta}'}^{2}}=&& \dfrac{2}{3}c_{\theta}[c_{\theta}(2m_{K}^{2}+
m_{\pi}^{2})+2\sqrt{2}(-m_{K}^{2}+m_{\pi}^{2})s_{\theta}]\Lambda_{2}
\nn \\ &&
+ \dfrac{16 L_{8}}{3F^{2}}[c_{\theta}^{2}(4m_{K}^{4}-4m_{K}^{2}m_{\pi}^{2}+3m_{\pi}^{4})+8\sqrt{2}c_{\theta}
m_{K}^{2}(-m_{K}^{2}+m_{\pi}^{2})s_{\theta} +(8m_{K}^{4}-8m_{K}^{2}m_{\pi}^{2}+3m_{\pi}^{4})s_{\theta}^{2}]
\nn \\ && 
+ \dfrac{1}{16\pi^{2}} \bigg\{ \dfrac{(4m_{K}^{2}-m_{\pi}^{2})(2c_{\theta}^{4}-2\sqrt{2}
c_{\theta}^{3}s_{\theta}-3c_{\theta}^{2}s_{\theta}^{2}+2\sqrt{2}c_{\theta}s_{\theta}^{3}+2s_{\theta}^{4})
}{18F^{2}} A_{0}(m_{\eta}^{2})\nn \\&&
+\dfrac{1}{18F^{2}} \bigg[2c_{\theta}^{4}(2m_{K}^{2}+m_{\pi}^{2})-16\sqrt{2}c_{\theta}^{3}(m_{K}^{2}-m_{\pi}^{2})
s_{\theta}+12c_{\theta}^{2}(4m_{K}^{2}-m_{\pi}^{2})s_{\theta}^{2}\nn \\&&
\quad -4\sqrt{2}c_{\theta}(8m_{K}^{2}-5m_{\pi}^{2})s_{\theta}^{3}+(16m_{K}^{2}-7m_{\pi}^{2})s_{\theta}^{4}\bigg]A_{0}(m_{\eta'}^{2})\nn \\&&
-\dfrac{[-4c_{\theta}^{2}m_{K}^{2}+2\sqrt{2}c_{\theta}(2m_{K}^{2}-m_{\pi}^{2})s_{\theta}+m_{\pi}^{2}s_{\theta}^{2}]}
{3F^{2}} A_{0}(m_{K}^{2})+\dfrac{m_{\pi}^{2}(2c_{\theta}^{2}+2\sqrt{2}c_{\theta}s_{\theta}+s_{\theta}^{2})}
{2F^{2}} A_{0}(m_{\pi}^{2}) \bigg\}
\nn \\&&
-\dfrac{16 L_{25}c_{\theta}[c_{\theta}(4m_{K}^{4}-4m_{K}^{2}m_{\pi}^{2}+3m_{\pi}^{4})+4\sqrt{2}m_{K}^{2}
(-m_{K}^{2}+m_{\pi}^{2})s_{\theta}]}{F^{2}} +6c_{\theta}^{2}(2m_{K}^{2}+m_{\pi}^{2})v_2^{(2)} \nn \\&&
+\dfrac{16 L_{7}[c_{\theta}^{2}(2m_{K}^{2}+m_{\pi}^{2})^{2}+4\sqrt{2}c_{\theta}(-2m_{K}^{4}+m_{K}^{2}m_{\pi}^{2}
+m_{\pi}^{4})s_{\theta}+8(m_{K}^{2}-m_{\pi}^{2})^{2}s_{\theta}^{2}]}{3F^{2}}\nn \\&&
+\dfrac{16 L_{6}(2m_{K}^2+m_{\pi}^{2})[c_{\theta}^{2}(2m_{K}^{2}+m_{\pi}^{2})+4\sqrt{2}c_{\theta}(-m_{K}^{2}+
m_{\pi}^{2})s_{\theta}+(4m_{K}^{2}-m_{\pi}^{2})s_{\theta}^{2}]}{3F^{2}}\nn \\&&
+\dfrac{256( L_{5}-2 L_{8}) L_{8}}{3F^{4}}[c_\theta^{2}(4m_{K}^{6}-2m_{K}^{4}m_{\pi}^{2}-2m_{K}^{2}m_{\pi}^{4}+3m_{\pi}^{6})+\nn \\&&
4\sqrt{2}c_\theta
m_{K}^{2}(-2m_{K}^{4}+m_{K}^{2}m_{\pi}^{2}+m_{\pi}^{4})s_\theta+(8m_{K}^{6}-4m_{K}^{4}m_{\pi}^{2}-4m_{K}^{2}m_{\pi}^{4}+3m_{\pi}^{6})s_\theta^{2}]
\nn \\&&
+\dfrac{16( L_{5}-2 L_{8})\Lambda_2 c_\theta[2\sqrt{2}s_\theta(-m_{K}^{4}+m_{\pi}^{4})+(2m_{K}^{4}+m_{\pi}^{4})c_\theta]}{3F^{2}}\nn \\&&
+\dfrac{16(3C_{19}+2C_{31})}{3F^{2}}\bigg[c_{\theta}^{2}(8m_{K}^{6}-12m_{K}^{4}m_{\pi}^{2}+6m_{K}^{2}m_{\pi}^{4} +m_{\pi}^{6} )
\nn \\&&
-4\sqrt{2}c_{\theta}(4m_{K}^{6}-6m_{K}^{4}m_{\pi}^{2}+3m_{K}^{2}m_{\pi}^{4}-m_{\pi}^{6})
s_{\theta} +(16m_{K}^{6}-24m_{K}^{4}m_{\pi}^{2}+12m_{K}^{2}m_{\pi}^{4}-m_{\pi}^{6})s_{\theta}^{2}\bigg] \,,
\end{eqnarray}
\begin{eqnarray}
 \delta_{m^{2}}=&& -\dfrac{64 L_{8}m_{K}^{2}(m_{K}^{2}-m_{\pi}^{2})(\sqrt{2}c_{\theta}^{2}-c_{\theta}s_{\theta}-\sqrt{2}
s_{\theta}^{2})}{3F^{2}}\nn \\&&
-\dfrac{2}{3}[\sqrt{2}c_{\theta}^{2}(m_{K}^{2}-m_{\pi}^{2})+c_{\theta}
(2m_{K}^{2}+m_{\pi}^{2})s_{\theta}+\sqrt{2}(-m_{K}^{2}+m_{\pi}^{2})s_{\theta}^{2}]\Lambda_{2}
\nn \\ &&
-\dfrac{1}{288\pi^{2} F^2}\bigg\{\,\, \bigg[ \sqrt{2}c_{\theta}^{4}(8m_{K}^{2}
-5m_{\pi}^{2})+c_{\theta}^{3}(8m_{K}^{2}+m_{\pi}^{2})s_{\theta}+3\sqrt{2}c_{\theta}^{2}(-4m_{K}^{2}
+m_{\pi}^{2})s_{\theta}^{2}
\nn \\&& \quad +4c_{\theta}(-5m_{K}^{2}+2m_{\pi}^{2})s_{\theta}^{3}+4\sqrt{2}(-m_{K}^{2}+m_{\pi}^{2})s_{\theta}^{4} \bigg] A_{0}(m_{\eta}^{2})
\nn \\&&
+ \bigg[4\sqrt{2}c_{\theta}^{4}(m_{K}^{2}-m_{\pi}^{2})+ 3\sqrt{2}c_{\theta}^{2}(4m_{K}^{2}-m_{\pi}^{2})s_{\theta}^{2}+c_{\theta}(8m_{K}^{2}+m_{\pi}^{2})s_{\theta}^{3}
+\sqrt{2}(-8m_{K}^{2}+5m_{\pi}^{2})s_{\theta}^{4}
\nn \\&& \quad +4c_{\theta}^{3}s_\theta(-5m_{K}^{2}+2m_{\pi}^{2}) \bigg]A_{0}(m_{\eta'}^{2})
\nn \\ && +6[\sqrt{2}c_{\theta}^{2}(2m_{K}^{2}-m_{\pi}^{2})+c_{\theta}(4m_{K}^{2}+m_{\pi}^{2})s_{\theta}
+\sqrt{2}(-2m_{K}^{2}+m_{\pi}^{2})s_{\theta}^{2}]A_{0}(m_{K}^{2})
\nn \\ && +9m_{\pi}^{2}(-\sqrt{2}c_{\theta}^{2}+c_{\theta}s_{\theta}+\sqrt{2}s_{\theta}^{2})A_{0}(m_{\pi}^{2}) \bigg\}
\nn \\&&
-6c_{\theta}(2m_{K}^{2}+m_{\pi}^{2})s_{\theta}v_2^{(2)}-\dfrac{32 L_{6}(2m_{K}^{4}-m_{K}^{2}m_{\pi}^{2}
-m_{\pi}^{4})(\sqrt{2}c_{\theta}^{2}-c_{\theta}s_{\theta}-\sqrt{2}s_{\theta}^{2})}{3F^{2}}\nn \\&&
+\dfrac{16 L_{25}}{F^{2}} \bigg[ 2\sqrt{2}c_{\theta}^{2}m_{K}^{2}(m_{K}^{2}-m_{\pi}^{2})+c_{\theta}(
4m_{K}^{4}-4m_{K}^{2}m_{\pi}^{2}+3m_{\pi}^{4})s_{\theta}+ 2\sqrt{2}m_{K}^{2}(-m_{K}^{2}+m_{\pi}^{2})s_{\theta}^{2} \bigg]\nn \\&&
-\dfrac{16 L_{7}}{3F^{2}} \bigg[2\sqrt{2}c_{\theta}^{2}
(2m_{K}^{4}-m_{K}^{2}m_{\pi}^{2}-m_{\pi}^{4})+c_{\theta}(-4m_{K}^{4}+20m_{K}^{2}m_{\pi}^{2}-7m_{\pi}^{4})s_{\theta}
+\nn \\ && \quad
2\sqrt{2}(-2m_{K}^{4}+m_{K}^{2}m_{\pi}^{2}
+m_{\pi}^{4})s_{\theta}^{2} \bigg]\nn \\&&
- \dfrac{512( L_{5}-2 L_{8}) L_{8}m_{K}^{2}(2m_{K}^{4}-m_{K}^{2}m_{\pi}^{2}-m_{\pi}^{4})(\sqrt{2}c_\theta^{2}-c_\theta
s_\theta-\sqrt{2}s_\theta^{2})}{3F^{4}}\nn \\&&
-\dfrac{16( L_{5}-2 L_{8})\Lambda_2[\sqrt{2}c_\theta^{2}(m_{K}^{4}-m_{\pi}^{4})+c_\theta(2m_{K}^{4}+m_{\pi}^{4})s_\theta+
\sqrt{2}(-m_{K}^{4}+m_{\pi}^{4})s_\theta^{2}]}{3F^{2}}\nn \\&&
-\dfrac{32(3C_{19}+2C_{31})(4m_{K}^{6}-6m_{K}^{4}m_{\pi}^{2}+3m_{K}^{2}m_{\pi}^{4}-m_{\pi}^{6})(\sqrt{2}
c_{\theta}^{2}-c_{\theta}s_{\theta}-\sqrt{2}s_{\theta}^{2})}{3F^{2}} \,.
\end{eqnarray}

When expressing the above results in terms of $F_\pi$ from Eq.~\eqref{fpinlof0}, the terms with $ L_5^2$ and $ L_5 L_8$
can be different from the expressions in terms of $F$ and the other parts remain the same, apart from the obvious replacement of $F$ by $F_\pi$.
Therefore, for the expressions of $\delta_i$ expressed in $F_\pi$, we only give the parts that are different from those in terms of $F$
\begin{eqnarray}
 \delta_{\overline{\eta}}^{(F_\pi), L_5^2}=
\dfrac{128 L_{5}^2 }{3F_\pi^{4}} [ c_\theta^{2}(2m_{K}^{4}+ 2m_K^2m_{\pi}^2- m_{\pi}^{4})
+2\sqrt{2}c_\theta s_\theta ( m_{K}^{4}+m_K^2m_{\pi}^2- 2m_{\pi}^{4})
+ s_\theta^{2} (m_{K}^{4}+ m_K^2m_{\pi}^2+ m_{\pi}^{4}) ] \,,
\end{eqnarray}
\begin{eqnarray}
 \delta_{\overline{\eta}'}^{(F_\pi), L_5^2}=
\dfrac{128 L_{5}^2 }{3F_\pi^{4}} [ c_\theta^{2}(m_{K}^{4}+ m_K^2m_{\pi}^2+ m_{\pi}^{4})
-2\sqrt{2}c_\theta s_\theta ( m_{K}^{4} + m_K^2m_{\pi}^2- 2m_{\pi}^{4})
+  s_\theta^{2} (2m_{K}^{4}+ 2m_K^2m_{\pi}^2- m_{\pi}^{4})  ] \,,
\end{eqnarray}
\begin{eqnarray}
 \delta_{k}^{(F_\pi), L_5^2}= -\dfrac{128 L_{5}^2 (m_{K}^{4}+m_K^2m_\pi^2-2m_{\pi}^{4})(\sqrt{2}c_\theta^{2}-c_\theta s_\theta-\sqrt{2}s_\theta^{2})}{3F_\pi^{4}}\,,
\end{eqnarray}
\begin{eqnarray}
 \delta_{m_{\overline{\eta}}^{2}}^{(F_\pi), L_5 L_8}= && \dfrac{128 L_{5} L_{8}}{3F_\pi^{4}}
\bigg[c_\theta^{2}(16m_{K}^{6}-16m_{K}^{2}m_{\pi}^{4}+9m_{\pi}^{6})+  16\sqrt{2}c_\theta
m_{K}^{2}(m_{K}^{4}-m_{\pi}^{4})s_\theta
\nn \\ && \qquad
+(8m_{K}^{6}-8m_{K}^{2}m_{\pi}^{4}+9m_{\pi}^{6})s_\theta^{2} \bigg]\,,
\end{eqnarray}
\begin{eqnarray}
 \delta_{m_{\overline{\eta}'}^{2}}^{(F_\pi), L_5 L_8}=&&  \dfrac{128 L_{5} L_{8}}{3F_\pi^{4}}
\bigg[c_\theta^{2}(8m_{K}^{6}-8m_{K}^{2}m_{\pi}^{4}+9m_{\pi}^{6}) -
16\sqrt{2}c_\theta m_{K}^{2}(m_{K}^{4}-m_{\pi}^{4})s_\theta
\nn \\ && \qquad
+(16m_{K}^{6}-16m_{K}^{2}m_{\pi}^{4}+9m_{\pi}^{6})s_\theta^{2} \bigg]\,,
\end{eqnarray}
\begin{eqnarray}
 \delta_{m^{2}}^{(F_\pi), L_5 L_8}= -\dfrac{1024 L_{5} L_{8}m_K^2 (m_K^4-m_\pi^4) (\sqrt2 c_\theta^2 -c_\theta s_\theta - \sqrt2 s_\theta^2) }{3F_\pi^{4}} \,.
\end{eqnarray}
In order to obtain the full expressions for the $\delta_i$'s given in terms of $F_\pi$ one has to
make use of Eq.~\eqref{fpinlof0} up to the precision required.
Taking $\delta_k$ for example, its final expression in terms of $F_\pi$ is
\begin{eqnarray}
 \delta_{k}=&& -\dfrac{16 L_{5}(m_{K}^{2}-m_{\pi}^{2})
(\sqrt{2}c_{\theta}^{2}-c_{\theta}s_{\theta}-\sqrt{2}s_{\theta}^{2})}{3F_\pi^{2}}-c_{\theta}s_{\theta}\Lambda_{1}\nn \\
&&+\dfrac{c_{\theta}s_{\theta}A_{0}(m_{K}^{2})}{16\pi^{2}F_\pi^{2}}
-\dfrac{8 L_{18}[\sqrt{2}c_{\theta}^{2}(m_{K}^{2}-m_{\pi}^{2})+c_{\theta}(2m_{K}^{2}+m_{\pi}^{2})s_{\theta}+\sqrt{2}
(-m_{K}^{2}+m_{\pi}^{2})s_{\theta}^{2}]}{F_\pi^{2}} \nn \\
&& +\dfrac{256 L_{5} L_{8}(m_{K}^{4}-m_{\pi}^{4})(\sqrt{2}c_\theta^{2}-c_\theta s_\theta-\sqrt{2}s_\theta^{2})}{3F_\pi^{4}}
-\dfrac{128 L_{5}^2 (m_{K}^{4}+m_K^2m_\pi^2-2m_{\pi}^{4})(\sqrt{2}c_\theta^{2}-c_\theta s_\theta-\sqrt{2}s_\theta^{2})}{3F_\pi^{4}} \nn \\
&& -\dfrac{64(C_{14}+C_{17})m_{K}^{2}(m_{K}^{2}-m_{\pi}^{2})(\sqrt{2}c_{\theta}^{2}-c_{\theta}s_{\theta}-\sqrt{2}s_{\theta}^{2})}
{3F_\pi^{2}} \,,
\end{eqnarray}
which differs from Eq.~\eqref{deltaknnlof0} in the $ L_5^2$ term.
For $\delta_1$, $\delta_2$ and $\delta_3$, their expressions are the same regardless
of whether $F$ or $F_\pi$ is chosen up to next-to-next-to-leading order.

For completeness, we also give the results in terms of the LO masses $\overline{m}_\pi$ and $\overline{m}_K$ and $1/F^2$. 
Only the terms with $L_i L_j$, being $L_i$ and $L_j$ the NLO LECs in Eq.~\eqref{lagnlo}, will be different, comparing with 
the expressions in terms of $m_\pi$ and $m_K$ and the other parts remain the same, apart from the obvious replacement of the renormalized 
masses by the LO ones. Therefore, we only give the parts that are different from those in terms of $m_\pi$, $m_K$ and $1/F^2$ and 
it turns out that in this case all of the $L_i L_j$ terms for 
$\delta_{\overline{\eta}}, \delta_{\overline{\eta}'}, \delta_{k}, \delta_{m_{\overline{\eta}}^{2}}, \delta_{m_{\overline{\eta}'}^{2}}, \delta_{m^{2}}$ 
vanish.


\begin{thebibliography}{90}
\bibitem{gasser8485}J.~Gasser and H.~Leutwyler, Ann. Phys. {\bf 158} (1984) 142; Nucl. Phys. B {\bf 250} (1985) 465.
\bibitem{Nc}
%\bibitem{'tHooft:1973jz}
  G.~'t Hooft,
  %``A Planar Diagram Theory for Strong Interactions,''
  Nucl.\ Phys.\ B {\bf 72} (1974) 461;
  %%CITATION = NUPHA,B72,461;%%
%
%\bibitem{'tHooft:1974hx}
%  G.~'t Hooft,
  %``A Two-Dimensional Model for Mesons,''
%  Nucl.\ Phys.\ B
    {\bf 75} (1974) 461;
  %%CITATION = NUPHA,B75,461;%%
%
%\bibitem{Witten:1979kh}
  E.~Witten,
  %``Baryons in the 1/n Expansion,''
  Nucl.\ Phys.\ B {\bf 160} (1979) 57.
  %%CITATION = NUPHA,B160,57;%%
%\cite{DescotesGenon:1999uh}
\bibitem{u3lo} P.~Di~Vecchia and G.~Veneziano, Nucl. Phys. B {\bf 171} (1980) 253;
C. Rosenzweig, J. Schechter and T. Trahem, Phys. Rev. D {\bf 21} (1980) 3388;
E.~Witten, Ann. Phys. {\bf 128} (1980) 363;  K.~Kawarabayashi and N.~Ohta,
  %``The Problem of $\eta$ in the Large $N$ Limit: Effective Lagrangian Approach,''
  Nucl.\ Phys.\ B {\bf 175}, 477 (1980).
  %%CITATION = NUPHA,B175,477;%%
  %191 citations counted in INSPIRE as of 24 Apr 2015
\bibitem{leutwyler98npbsp}
  H.~Leutwyler,
  %``On the 1/N expansion in chiral perturbation theory,''
  Nucl.\ Phys.\ Proc.\ Suppl.\  {\bf 64}, 223 (1998).
\bibitem{Kaiser:1998ds}
  R.~Kaiser and H.~Leutwyler, hep-ph/9806336.
\bibitem{kaiser00epjc}R.~Kaiser and H.~Leutwyler, Eur.~Phys.~J.~{\bf C17} (2000)623.
\bibitem{herrera97npb}P.~Herrera-Siklody, J.~I.~Latorre, P.~Pascual and J.~Taron, Nucl.~Phys.~{\bf B497} (1997) 345.
\bibitem{herrera98plb}P.~Herrera-Siklody, J.~I.~Latorre, P.~Pascual and J.~Taron, Phys.~Lett.~{\bf B419} (1998) 326.
\bibitem{hsc11prd}
  J.~J.~Dudek, R.~G.~Edwards, B.~Joo, M.~J.~Peardon, D.~G.~Richards and C.~E.~Thomas,
  %``Isoscalar meson spectroscopy from lattice QCD,''
  Phys.\ Rev.\ D {\bf 83}, 111502 (2011).
%  [arXiv:1102.4299 [hep-lat]].
  %%CITATION = ARXIV:1102.4299;%%
  %90 citations counted in INSPIRE as of 17 Dec 2014
\bibitem{rbcukqcd10prl}
  N.~H.~Christ, C.~Dawson, T.~Izubuchi, C.~Jung, Q.~Liu, R.~D.~Mawhinney, C.~T.~Sachrajda and A.~Soni {\it et al.},
  %``The $\eta$ and $\eta^\prime$ mesons from Lattice QCD,''
  Phys.\ Rev.\ Lett.\  {\bf 105}, 241601 (2010).
%  [arXiv:1002.2999 [hep-lat]].
  %%CITATION = ARXIV:1002.2999;%%
  %48 citations counted in INSPIRE as of 17 Dec 2014
%\cite{Dudek:2011tt}
\bibitem{ukqcd12prd}
  E.~B.~Gregory {\it et al.}  [UKQCD Collaboration],
  %``A study of the eta and eta' mesons with improved staggered fermions,''
  Phys.\ Rev.\ D {\bf 86}, 014504 (2012).
%  [arXiv:1112.4384 [hep-lat]].
  %%CITATION = ARXIV:1112.4384;%%
  %22 citations counted in INSPIRE as of 17 Dec 2014
%\cite{Christ:2010dd}
\bibitem{etm13pos}
  C.~Michael {\it et al.}  [European Twisted Mass Collaboration],
  %``$\eta$ and $\eta'$ masses and decay constants from lattice QCD with  $N_f=2+1+1$ quark flavours,''
  PoS LATTICE {\bf 2013}, 253 (2014).
%  [arXiv:1311.5490 [hep-lat]].
  %%CITATION = ARXIV:1311.5490;%%
  %5 citations counted in INSPIRE as of 17 Dec 2014
%\cite{Gregory:2011sg}
\bibitem{etm13prl}
  C.~Michael {\it et al.}  [ETM Collaboration],
  %``$\eta$ and $\eta^\prime$ mixing from Lattice QCD,''
  Phys.\ Rev.\ Lett.\  {\bf 111}, no. 18, 181602 (2013).
%  [arXiv:1310.1207 [hep-lat]].
  %%CITATION = ARXIV:1310.1207;%%
  %16 citations counted in INSPIRE as of 17 Dec 2014
%\cite{Michael:2013vba}
\bibitem{chen12prd}
  Y.~H.~Chen, Z.~H.~Guo and H.~Q.~Zheng,
  %``Study of \eta-\eta' mixing from radiative decay processes,''
  Phys.\ Rev.\ D {\bf 85}, 054018 (2012).
\bibitem{Chen:2014yta}
  Y.~H.~Chen, Z.~H.~Guo and B.~S.~Zou,
  %``Unified study of $J/\psi \to PV$, $P\gamma^{(*)}$ and light hadron radiative processes,''
  Phys.\ Rev.\ D {\bf 91}, 014010 (2015).
%  [arXiv:1411.1159 [hep-ph]].
  %%CITATION = ARXIV:1411.1159;%%
  %1 citations counted in INSPIRE as of 17 Feb 2015
%\cite{Aoki:2013ldr}

\bibitem{rbcukqcd11prd}
  Y.~Aoki {\it et al.}  [RBC and UKQCD Collaborations],
  %``Continuum Limit Physics from 2+1 Flavor Domain Wall QCD,''
  Phys.\ Rev.\ D {\bf 83} (2011) 074508.
%  [arXiv:1011.0892 [hep-lat]].
  %%CITATION = ARXIV:1011.0892;%%
  %109 citations counted in INSPIRE as of 04 Dec 2013
%\cite{Arthur:2012opa}
\bibitem{rbcukqcd13prd}
  R.~Arthur {\it et al.}  [RBC and UKQCD Collaborations],
  %``Domain Wall QCD with Near-Physical Pions,''
  Phys.\ Rev.\ D {\bf 87} (2013) 094514.
%  [arXiv:1208.4412 [hep-lat]].
  %%CITATION = ARXIV:1208.4412;%%
  %33 citations counted in INSPIRE as of 04 Dec 2013
\bibitem{Durr:2010hr}
%%%``The ratio FK/Fpi in QCD,''
    S.~Durr {\it et al.},
%%%, Z.~Fodor, C.~Hoelbling, S.~D.~Katz, S.~Krieg, T.~Kurth, L.~Lellouch and T.~Lippert {\it et al.},
  Phys.\ Rev.\ D {\bf 81} (2010) 054507.
%  [arXiv:1001.4692 [hep-lat]].
%\cite{Bijnens:2014lea}
\bibitem{DescotesGenon:1999uh}
  S.~Descotes-Genon, L.~Girlanda and J.~Stern,
  %``Paramagnetic effect of light quark loops on chiral symmetry breaking,''
  JHEP {\bf 0001}, 041 (2000).
%  [hep-ph/9910537].
  %%CITATION = HEP-PH/9910537;%%
  %89 citations counted in INSPIRE as of 26 Feb 2015
%\cite{Amoros:1999dp}
\bibitem{Bijnens:2011tb}
  J.~Bijnens and I.~Jemos,
  %``A new global fit of the $L^r_i$ at next-to-next-to-leading order in Chiral Perturbation Theory,''
  Nucl.\ Phys.\ B {\bf 854}, 631 (2012).
%  [arXiv:1103.5945 [hep-ph]].
  %%CITATION = ARXIV:1103.5945;%%
  %46 citations counted in INSPIRE as of 29 Jan 2015
\bibitem{Bijnens:2014lea}
  J.~Bijnens and G.~Ecker,
  %``Mesonic low-energy constants,''
  Ann.\ Rev.\ Nucl.\ Part.\ Sci.\  {\bf 64}, 149 (2014).
%  [arXiv:1405.6488 [hep-ph]].
  %%CITATION = ARXIV:1405.6488;%%
  %11 citations counted in INSPIRE as of 29 Jan 2015
%\cite{Jiang:2009uf}
\bibitem{Ecker:2013pba}
  G.~Ecker, P.~Masjuan and H.~Neufeld,
  %``Approximating chiral $SU(3)$ amplitudes,''
  Eur.\ Phys.\ J.\ C {\bf 74}, no. 2, 2748 (2014).
%  [arXiv:1310.8452 [hep-ph]].
  %%CITATION = ARXIV:1310.8452;%%
  %7 citations counted in INSPIRE as of 17 Feb 2015

%\cite{Guo:2014yva}
\bibitem{Guo:2014yva}
  Z.~H.~Guo and J.~J.~Sanz-Cillero,
  %``Resonance effects in pion and kaon decay constants,''
  Phys.\ Rev.\ D {\bf 89}, no. 9, 094024 (2014).
%  [arXiv:1403.0855 [hep-ph]].
  %%CITATION = ARXIV:1403.0855;%%
  %2 citations counted in INSPIRE as of 17 Feb 2015
%\cite{Ecker:2013pba}
\bibitem{feldmann98prd} T. Feldmann, P. Kroll and B. Stech, Phys.~Rev.~D~{\bf 58}, 114006 (1998).
\bibitem{escribano05jhep} R. Escribano and J. M. Frere, J. High Energy Phys. {\bf 0506}, 029 (2005).
\bibitem{thomas07jhep} C.~E.~Thomas, J.~High~Energy~Phys.~{\bf 0710}, 026 (2007).
\bibitem{zhao08jpg} G.~Li, Q.~Zhao, and C.~H.~Chang, J.\ Phys.\ G {\bf 35}, 055002 (2008).
\bibitem{escribano14prd}
  R.~Escribano, P.~Masjuan and P.~Sanchez-Puertas,
  %``$\eta$ and $\eta^{\prime}$ transition form factors from rational approximants,''
  Phys.\ Rev.\ D {\bf 89}, no. 3, 034014 (2014). 
\bibitem{DeFazio:2000my}
  F.~De Fazio and M.~R.~Pennington,
  %``Radiative phi meson decays and eta - eta-prime mixing: A QCD sum rule analysis,''
  JHEP {\bf 0007}, 051 (2000)
  [hep-ph/0006007].
  %%CITATION = HEP-PH/0006007;%% 
%\cite{Schechter:1992iz}
\bibitem{Schechter:1992iz} 
  J.~Schechter, A.~Subbaraman and H.~Weigel,
  %``Effective hadron dynamics: From meson masses to the proton spin puzzle,''
  Phys.\ Rev.\ D {\bf 48}, 339 (1993)
  [hep-ph/9211239].
  %%CITATION = HEP-PH/9211239;%%
  %116 citations counted in INSPIRE as of 24 Apr 2015
\bibitem{ua1nc} E.~Witten, Nucl.\ Phys.\ B {\bf 156}, 269 (1979); S.~Coleman and E.~Witten, Phys.~Rev.~Lett.~{\bf 45}, 100 (1980);
G.~Veneziano, Nucl.\ Phys.\ B {\bf 159}, 213 (1979).
\bibitem{Jiang:2014via}
  S.~Z.~Jiang, F.~J.~Ge and Q.~Wang,
  %``Full pseudoscalar mesonic chiral Lagrangian at p6 order under the unitary group,''
  Phys.\ Rev.\ D {\bf 89}, 074048 (2014).
\bibitem{borasoy01epja}
  N.~Beisert and B.~Borasoy,
  %``eta eta-prime mixing in U(3) chiral perturbation theory,''
  Eur.\ Phys.\ J.\ A {\bf 11}, 329 (2001).
%  [hep-ph/0107175].
  %%CITATION = HEP-PH/0107175;%%
  %25 citations counted in INSPIRE as of 26 Jan 2015
\bibitem{Gerard:2004gx}
  J.-M.~Gerard and E.~Kou,
  %``eta-eta-prime masses and mixing: A Large N(c) reappraisal,''
  Phys.\ Lett.\ B {\bf 616}, 85 (2005).
\bibitem{Gerard:2009ps}
  C.~Degrande and J.-M.~Gerard,
  %``A Theoretical determination of the eta - eta-prime mixing,''
  JHEP {\bf 0905}, 043 (2009).
\bibitem{Mathieu:2010ss}
  V.~Mathieu and V.~Vento,
  %``$\eta-\eta^\prime$ mixing in the flavor basis and large N,''
  Phys.\ Lett.\ B {\bf 688}, 314 (2010).
\bibitem{Guo:2011pa}
  Z.~-H.~Guo and J.~A.~Oller,
  %``Resonances from meson-meson scattering in U(3) CHPT,''
  Phys.\ Rev.\ D {\bf 84} (2011) 034005.
%  [arXiv:1104.2849 [hep-ph]].
  %%CITATION = ARXIV:1104.2849;%%
  %28 citations counted in INSPIRE as of 26 Sep 2013
 %\cite{Guo:2012yt}
\bibitem{bijnens99jhep} J.~Bijnens, G.~Colangelo and G.~Ecker, JHEP~{\bf 02} (1999) 020.
%\cite{Guo:2011pa}
\bibitem{Guo:2012yt}
  Z.~-H.~Guo, J.~A.~Oller and J.~Ruiz de Elvira,
  %``Chiral dynamics in form factors, spectral-function sum rules, meson-meson scattering and semi-local duality,''
  Phys.\ Rev.\ D {\bf 86} (2012) 054006.
  %%CITATION = ARXIV:1206.4163;%%
  %11 citations counted in INSPIRE as of 26 Sep 2013
%\cite{Amoros:2001cp}
\bibitem{Amoros:2001cp} 
  G.~Amoros, J.~Bijnens and P.~Talavera,
  %``QCD isospin breaking in meson masses, decay constants and quark mass ratios,''
  Nucl.\ Phys.\ B {\bf 602}, 87 (2001).
%  [hep-ph/0101127].
  %%CITATION = HEP-PH/0101127;%%
  %213 citations counted in INSPIRE as of 24 Apr 2015
  
\bibitem{Georgi:1993jn}
  H.~Georgi,
  %``A bound on m(eta) / m(eta-prime) for large n(c),''
  Phys.\ Rev.\ D {\bf 49}, 1666 (1994).
\bibitem{Peris:1993np}
  S.~Peris,
  %``Higher order corrections to the large N(c) bound on M(eta) / M(eta-prime),''
  Phys.\ Lett.\ B {\bf 324}, 442 (1994).
\bibitem{Feldmann:1999uf}
For a recent review:  T.~Feldmann,
  %``Quark structure of pseudoscalar mesons,''
  Int.\ J.\ Mod.\ Phys.\ A {\bf 15} (2000) 159.
%  [arXiv:hep-ph/9907491].
  %%CITATION = HEP-PH/9907491;%%
  %257 citations counted in INSPIRE as of 23 Nov 2013
%\cite{Michael:2013gka}
\bibitem{Amoros:1999dp}
  G.~Amoros, J.~Bijnens and P.~Talavera,
  %``Two point functions at two loops in three flavor chiral perturbation theory,''
  Nucl.\ Phys.\ B {\bf 568}, 319 (2000).
%  [hep-ph/9907264].
  %%CITATION = HEP-PH/9907264;%%
  %129 citations counted in INSPIRE as of 28 Feb 2015
\bibitem{Bernard:2009ds}
  V.~Bernard and E.~Passemar,
  %``Chiral Extrapolation of the Strangeness Changing K pi Form Factor,''
  JHEP {\bf 1004}, 001 (2010).
%  [arXiv:0912.3792 [hep-ph]].
  %%CITATION = ARXIV:0912.3792;%%
  %23 citations counted in INSPIRE as of 25 Dec 2014kda
%\cite{Beisert:2001qb}
\bibitem{Aoki:2013ldr}
  S.~Aoki, Y.~Aoki, C.~Bernard, T.~Blum, G.~Colangelo, M.~Della Morte, S.~Durr and A.~X.~El Khadra {\it et al.},
  %``Review of lattice results concerning low-energy particle physics,''
  Eur.\ Phys.\ J.\ C {\bf 74}, no. 9, 2890 (2014).
%  [arXiv:1310.8555 [hep-lat]].
  %%CITATION = ARXIV:1310.8555;%%
  %160 citations counted in INSPIRE as of 17 Feb 2015

%\cite{Agashe:2014kda}
\bibitem{pdg14}
  K.~A.~Olive {\it et al.}  [Particle Data Group Collaboration],
  %``Review of Particle Physics,''
  Chin.\ Phys.\ C {\bf 38}, 090001 (2014).
  %%CITATION = CHPHD,C38,090001;%%
  %348 citations counted in INSPIRE as of 19 Dec 2014
%\cite{Bernard:2009ds}

%\cite{Escribano:2010wt}
\bibitem{Escribano:2010wt}
  R.~Escribano, P.~Masjuan and J.~J.~Sanz-Cillero,
  %``Chiral dynamics predictions for eta' -> eta pi pi,''
  JHEP {\bf 1105}, 094 (2011).
%  [arXiv:1011.5884 [hep-ph]].
  %%CITATION = ARXIV:1011.5884;%%
  %23 citations counted in INSPIRE as of 16 Feb 2015

\bibitem{Ecker89}G. Ecker {\it et al.}, Nucl. Phys. {\bf B321}, 311 (1989).
\bibitem{L8-rcht}
%\bibitem{Rosell-L8}
%%%Towards a determination of the chiral couplings at NLO in 1/N(C): L**r(8)(mu)
    A. Pich,  I. Rosell and J.J. Sanz-Cillero,
    JHEP {\bf 0701} (2007) 039.
%    [arXiv:hep-ph/0610290];
%
  %\cite{SanzCillero:2009ap}
%\bibitem{SanzCillero:2009ap}
  J.~J.~Sanz-Cillero and J.~Trnka,
  %``High energy constraints in the octet SS-PP correlator and resonance saturation at NLO in 1/N(c),''
  Phys.\ Rev.\ D {\bf 81} (2010) 056005.
%  [arXiv:0912.0495 [hep-ph]].
  %%CITATION = ARXIV:0912.0495;%%
  %10 citations counted in INSPIRE as of 14 Nov 2013
%\cite{Pich:2010sm}

%\cite{Jamin:2000wn}
\bibitem{Jamin:2000wn}
  M.~Jamin, J.~A.~Oller and A.~Pich,
  %``S wave K pi scattering in chiral perturbation theory with resonances,''
  Nucl.\ Phys.\ B {\bf 587}, 331 (2000).
%  [hep-ph/0006045].
  %%CITATION = HEP-PH/0006045;%%
  %177 citations counted in INSPIRE as of 03 Mar 2015
\bibitem{Jiang:2009uf}
  S.~Z.~Jiang, Y.~Zhang, C.~Li and Q.~Wang,
  %``Computation of the p**6 order chiral Lagrangian coefficients,''
  Phys.\ Rev.\ D {\bf 81}, 014001 (2010).
%  [arXiv:0907.5229 [hep-ph]].
  %%CITATION = ARXIV:0907.5229;%%
  %11 citations counted in INSPIRE as of 29 Jan 2015
%\cite{Bijnens:2011tb}

\bibitem{Jiang15}   S.~Z.~Jiang, Z.~L.~Wei, Q.~S.~Chen and Q.~Wang,
  %``Computation of the O(p^6) order low-energy constants: an update,''
  arXiv:1502.05087 [hep-ph].
  %%CITATION = ARXIV:1502.05087;%%
%\cite{Chen:2014yta}

%\cite{Oller:2006xb}
\bibitem{Oller:2006xb} 
  J.~A.~Oller and L.~Roca,
  %``Non-Perturbative Study of the Light Pseudoscalar Masses in Chiral Dynamics,''
  Eur.\ Phys.\ J.\ A {\bf 34}, 371 (2007). 
%  [hep-ph/0608290].
  %%CITATION = HEP-PH/0608290;%%
  %5 citations counted in INSPIRE as of 19 May 2015


\end{thebibliography}
\end{document}